\documentclass[10pt,prx,aps,twocolumn,superscriptaddress]{revtex4-2}

\usepackage[usenames, dvipsnames]{xcolor}
\usepackage[utf8]{inputenc}
\usepackage{hyperref}
\hypersetup{colorlinks,allcolors=blue}
\usepackage{amsmath,revsymb,amssymb,mathtools}
\usepackage{graphicx}
\usepackage{xcolor}
\usepackage{bbold,comment}
\usepackage{braket,ulem}
\usepackage{booktabs}
\usepackage{float}
\usepackage{physics}
\usepackage{parskip}
\usepackage{multirow}
\usepackage{tabularx}
\usepackage{array}
\usepackage{pgf}
\usepackage{csquotes}

\newcolumntype{C}[1]{>{\centering\let\newline\\\arraybackslash\hspace{0pt}}m{#1}}
\newcolumntype{Y}{>{\centering\arraybackslash}X}

\renewcommand{\emph}{\textit}

\newtheorem{result}{Result}
\newcommand{\aliceobs}[1]{A_{#1}}
\newcommand{\aliceproj}[2]{\Lambda^{#1}_{#2}}

\newcommand{\bobobs}[1]{B_{#1}}
\newcommand{\bobproj}[2]{\Pi^{#1}_{#2}}

\newcommand{\til}{~}

\begin{document}
    
\title{Quantum bounds and device-independent security with rank-one qubit measurements}

\author{Lorenzo Coccia}
\email{lorenzo.coccia@unipd.it}
\affiliation{Dipartimento di Ingegneria dell'Informazione, Universit\`a degli Studi di Padova, via Gradenigo 6B, IT-35131 Padova, Italy}

\author{Matteo Padovan}
\affiliation{Dipartimento di Ingegneria dell'Informazione, Universit\`a degli Studi di Padova, via Gradenigo 6B, IT-35131 Padova, Italy}

\author{Andrea Pompermaier}
\affiliation{Dipartimento di Ingegneria dell'Informazione, Universit\`a degli Studi di Padova, via Gradenigo 6B, IT-35131 Padova, Italy}

\author{Mattia Sabatini}
\affiliation{Dipartimento di Ingegneria dell'Informazione, Universit\`a degli Studi di Padova, via Gradenigo 6B, IT-35131 Padova, Italy}

\author{Marco Avesani}
\affiliation{Dipartimento di Ingegneria dell'Informazione, Universit\`a degli Studi di Padova, via Gradenigo 6B, IT-35131 Padova, Italy}
\affiliation{Padua Quantum Technologies Research Center, Universit\`a degli Studi di Padova, via Gradenigo 6B, IT-35131 Padova, Italy}

\author{Davide G. Marangon}
\affiliation{Dipartimento di Ingegneria dell'Informazione, Universit\`a degli Studi di Padova, via Gradenigo 6B, IT-35131 Padova, Italy}

\author{Paolo Villoresi}
\affiliation{Dipartimento di Ingegneria dell'Informazione, Universit\`a degli Studi di Padova, via Gradenigo 6B, IT-35131 Padova, Italy}
\affiliation{Padua Quantum Technologies Research Center, Universit\`a degli Studi di Padova, via Gradenigo 6B, IT-35131 Padova, Italy}

\author{Giuseppe Vallone}
\affiliation{Dipartimento di Ingegneria dell'Informazione, Universit\`a degli Studi di Padova, via Gradenigo 6B, IT-35131 Padova, Italy}
\affiliation{Padua Quantum Technologies Research Center, Universit\`a degli Studi di Padova, via Gradenigo 6B, IT-35131 Padova, Italy}

\begin{abstract}
    Device-independent (DI) quantum protocols exploit Bell inequality violations to ensure security or certify quantum properties without making assumptions about the internal workings of the devices.
    In this work, we study the role of rank-one qubit positive operator-valued measures (POVMs) in DI scenarios.
    This class includes all qubit extremal POVMs, i.e., those measurements that cannot be realized by randomly choosing among others, as well as part of non-extremal POVMs, which have recently been shown to be useful for security applications in sequential quantum protocols.
    We demonstrate that any rank-one POVM can generate correlations in bipartite scenarios that saturate a Tsirelson inequality, i.e., a quantum bound on linear combinations of outcome statistics, when the two parties share an arbitrary entangled two-qubit state and some other self-tested measurements are performed.
    For extremal POVMs, such saturation allows for an explicit calculation of the guessing probability and the worst-case conditional von Neumann entropy. 
    From the Tsirelson inequality, we establish a randomness certification method that facilitates numerical simulations and noise analysis. To test its feasibility, we performed a proof-of-concept experiment employing a three-outcome POVM on tilted entangled states under experimental non-idealities. 
    We further explore the case of non-extremal POVMs, providing insights into their role in DI protocols. 
\end{abstract}

\maketitle
\section{Introduction}
In recent years, the field of device-independent (DI) quantum information processing has seen significant advancements \cite{scarani2012device,Supic_2020, Zapatero2023,Primaatmaja2023securityofdevice}. 
This framework involves spatially separated parties performing repeated measurements on a shared quantum system and estimating the probability distribution of possible outcomes. 
Remarkably, even without prior knowledge of the system, its state, or the measurement settings, the security of a protocol can be certified from the experimental outcomes based solely on the validity of quantum mechanics.

In DI scenarios the geometry of observed correlations plays a fundamental role \cite{Brunner2014,
Christensen2015,Goh_2018, PhysRevLett.108.100402,Padovan:2023mxc}. 
A quantum advantage can be obtained when correlations between the parties violate a Bell inequality, demonstrating that they lie outside the set with a local-realistic description.
However, analytical DI security proofs typically require a Bell inequality to be \emph{maximally} violated, meaning that the correlations must lie on the boundary of the set $\mathcal{Q}$ of correlations obtainable within quantum mechanics.
Equivalently, this is expressed by stating that the correlations saturate a Tsirelson inequality, namely a bound on linear combinations of the correlations which characterize the limits of $\mathcal{Q}$ \cite{Cirelson1980quantum,Brunner2014}. 
Such saturation often enables some form of self-testing of the apparatus or significantly simplifies security proofs \cite{Supic_2020,mayers2004selftestingquantumapparatus}. 
Therefore, identifying the presence of a saturated Tsirelson bound is of great importance for studying and implementing device-independent protocols.

Our work primarily concentrates on device-independent protocols for randomness generation, although many of the insights may extend to other types of protocols.
We consider bipartite scenarios where the correlations are obtained from two spatially separated parties, Alice and Bob, by averaging over the results of sufficiently many independent and identically distributed trials.
The measurements from which we seek to obtain randomness are positive operator-valued measures (POVMs) composed of rank-one elements and acting on a qubit space.
An important subclass of these measurements are extremal POVMs, namely those POVMs that cannot be expressed as non-trivial convex combination of others \cite{D'Ariano_2005}. 
Extremal POVMs are the most relevant for DI randomness generation, as any non-extremal POVM can be realized with statistical mixtures of extremal POVMs.
However, non-extremal POVMs can also be beneficial in certain DI scenarios: 
One example occurs when the possible correlations are restricted not only by the no-signaling condition between the separated parties but also by constraints on the temporal sequence of local measurements \cite{Padovan:2023mxc}.

In typical protocols, along with the POVM whose correlations are used to extract randomness, it is necessary to consider a set of unitary measurements on Alice's and Bob's sides that can maximally violate a specific set of tilted CHSH inequalities \cite{Acin_2016, Woodhead_2020}, or other inequalities like the elegant Bell inequality \cite{Acin_2016, PhysRevA.97.012314, Tavakoli_2021}. 
Such violations enable a form of self-testing for both the shared state and the unitary measurements. 
The full joint correlations between the POVM and the self-tested measurements are then used to extract information about the POVM, in a process sometimes referred to as ``post-hoc" self-testing \cite{Supic_2020}. 
This approach has been employed to compute the guessing probability of an eavesdropper on the measurement outcomes in the case of extremal POVMs acting on qubits \cite{Acin_2016, Woodhead_2020}.
However, the guessing probability generally underestimate the available randomness in a protocol. 
A more accurate estimation should come from the worst-case conditional von Neumann entropy, optimized over all post-measurement states that are compatible with the experimental measurements \cite{renner2006securityquantumkeydistribution, Brown_2020, Devetak_2005}. 
Nevertheless, analytical general proofs for this von Neumann entropy, similar to those for the guessing probability, are currently absent. 
One technical difficulty is that the proofs for the guessing probability do not \emph{directly} rely on quantum boundary conditions involving the POVMs themselves, making generalization challenging.

The main result of this work is that with \emph{any} rank-one qubit POVM measured on \emph{any} entangled qubit state it is possible to obtain correlations, with a separated party, that lie on the boundary of the quantum set. 
We express this condition through the saturation of a Tsirelson inequality.
When extremal POVMs are used, this saturation implies the uniqueness of the correlations and the factorization of the post-measurement state with respect to any potential eavesdropper. 
With these results, the guessing probability  can be explicitly calculated, finding agreement with what already present in the literature. 
Moreover, the factorization between the user and the eavesdropper allows for the analytical determination of the worst-case conditional von Neumann entropy.
Notably, our randomness certification is not dependent on the entire joint correlations between the two separated parties, but solely on the condition of boundary saturation. 
Beyond the geometric significance, this simplification could facilitate numerical simulations and noise analysis. 
To demonstrate the practical feasibility of our certification method, we implemented a proof-of-principle experiment based on a photonic setup. 
The latter employs a non-projective 3-outcome POVM acting on tilted entangled states, and our analysis numerically accounts for experimental non-idealities.
Finally, we provide examples demonstrating how device-independent randomness might be bounded when non-extremal POVMs are considered, a case not previously addressed in the literature.
From a fundamental point of view, our methods can be used to further study the geometry of quantum correlations when POVMs are considered, or to further investigate the role of the different assumptions in the self-testing scenario \cite{baptista2023}.

The article is structured as follows. 
In Section \ref{sec:Methods}, we present the theoretical framework, in particular we establish the relationship between rank-one POVMs, qubit entangled states and the boundary of the quantum set of correlations. 
In Section \ref{sec:Randomness_Estimation}, we show how this relationship can be useful for randomness generation and certification. 
In Section \ref{sec:examples}, we apply these results to different DI-QRNG protocols implemented with various POVMs, both extremal and non-extremal.
In Section \ref{section:experimental_implementation}, we present the experimental implementation.
Section \ref{sec:Conclusions} concludes with a discussion of our results and future directions.

\section{Methods}
\label{sec:Methods}
We consider two spatially separated parties, Alice and Bob, performing a series of measurements on a bipartite system of two qubits.
Without losing generality, using the Schmidt decomposition,
the joint system can be written as
\begin{equation}
\label{eq:state_Bloch}
    \ket{\phi_{\theta}}=\cos \qty(\frac{\theta}{2}) \ket{00}+\sin \qty(\frac{\theta}{2}) \ket{11}\,,
\end{equation}
where $\{\ket0,\ket1\}$ is the qubit basis for both Alice and Bob and $\theta \in ]0,\pi[$. 
We aim to generate random numbers from the $d$ possible outcomes of a rank-one POVM $\{F_b\}$ ($b=0,\cdots,d-1$) performed by Bob, i.e.
\begin{equation}
\label{eq:F_proj}
F_b=k_b\ketbra{\beta_b}\,, \qquad \sum_{b=0}^{d-1} F_b=\openone\,,
\end{equation}
with $k_b > 0$ and $\ket{\beta_b}$ normalized vectors in the qubit space. 
By considering rank-one POVMs, we include all extremal POVMs in a bidimensional Hilbert space \cite{D'Ariano_2005}, which are those that cannot be expressed as convex combinations of other POVMs. 
However, there are rank-one POVMs which are non-extremal (see, for example, Sec.\ \ref{sec:non_ext_4}).

In the device-independent framework, the joint state \eqref{eq:state_Bloch} and the measurements \eqref{eq:F_proj} are unknown. 
In principle, the state shared between the two parties could even be mixed and entangled with an eavesdropper, Eve.
However, by considering its purification in a larger Hilbert space we can describe it as a tripartite pure state, 
$\ket{\psi} \in \mathcal{H}_A \otimes \mathcal{H}_B \otimes \mathcal{H}_E$, without losing in generality. 
Similarly, through the Stinespring dilation, the measurements on the state can be always modeled as orthogonal projective measurements \cite{Stinespring1955}. 
Specifically, we highlight that, given any measurement strategy (in general non-projective) performed by Alice and Bob, and generating some observed joint probability distribution, a projective strategy leading to the same distribution can always be found (see Theorem 3.13 and Lemma 3.15 in \cite{baptista2023}).
In particular, the measurements from which randomness is extracted are denoted by $\Pi_b$ acting on Bob's Hilbert space, $\mathcal{H}_B$ and  satisfying 
\begin{equation}
\sum_b \Pi_b=\openone \ , \qquad \Pi_b \Pi_{b'}=\delta_{bb'}\Pi_b \,.
\end{equation}

The standard way to certify DI randomness from POVM outcomes relies on self-testing results \cite{PhysRevLett.108.100402, Woodhead_2020,PhysRevA.91.052111,PhysRevA.98.042336}. 
Typically, it involves considering an extra set of unitary measurements on Alice's and Bob's sides, $A_j$ and $B_k$ ($j=1,2,3$ and $k=1, \dots , 6$), that can maximally violate an appropriate set of tilted CHSH inequalities. 
If such a violation is observed, it is possible to conclude that, up to local unitary transformations and possibly enlarging the Hilbert space (see Lemma 1 in \cite{Woodhead_2020}) the shared state between Alice and Bob is $\ket{\psi}=\ket{\phi_{\theta}}\ket{\xi}$, where $\ket{\phi_{\theta}}$ is the state \eqref{eq:state_Bloch} and $\ket{\xi}$ is an unknown state, possibly entangled with an eavesdropper Eve.
Alice's operators act trivially on $\ket{\xi}$ while on $\ket{\phi_\theta}$ their action is
\begin{equation}
\label{eq:op_alice}
\begin{split}
    A_1\ket{\psi}&= (\sigma_1\otimes \openone)\ket{\phi_{\theta}}\ket{\xi}  \ , \\ A_2\ket{\psi}&=(\sigma_2\otimes C)\ket{\phi_{\theta}}\ket{\xi}  \ , \\ A_3\ket{\psi}&=(\sigma_3\otimes \openone)\ket{\phi_{\theta}}\ket{\xi} \ .
\end{split}
\end{equation} 
The operator $C$ is a $\pm 1$ valued hermitian operator and it appears to take into account that correlations are invariant under complex conjugation, which cannot be described by a local unitary transformation \cite{mckague2010quantuminformationprocessingadversarial,McKague_2011,PhysRevA.98.042336}. 
While the precise form of $C$ is not directly needed for our analysis, the observed Bell inequality violation guarantees that Alice's operators anti-commute on the state \cite{Woodhead_2020,PhysRevA.98.042336}:
\begin{equation}
    \{A_i,A_j \}\ket{\psi}=0 \qquad i \neq j \ .
\end{equation}
and the joint correlations between Alice and Eve factorize:
\begin{equation}
\label{eq:fact_alice_eve}
    \expval{A_i E_e}=\expval{A_i}\expval{E_e}\,,
\end{equation}
with $E_e$ a generic operator of Eve.

After this first self-testing step, the problem can be expressed as follows: Alice and Bob have self-tested the action of Alice's operators $A_j$ \eqref{eq:op_alice} on the shared state $\ket{\phi_{\theta}}\ket\xi$. 
Our aim is to  generate random numbers from the $d$-outcome measurement $\qty{\Pi_b}$ satisfying
\begin{align}
\label{eq:constr_mean_value_0}
\expval{\openone \otimes \Pi_b}{\psi}
    &=\expval{\openone\otimes F_b}{\phi_\theta}\,, \\
\label{eq:constr_mean_value}
   \expval{A_j \otimes \Pi_b}{\psi}
    &=\expval{\sigma_j\otimes F_b}{\phi_\theta} \,.
\end{align}
These joint observed correlations allow to characterize the elements of the POVM on the space of $\ket{\phi_{\theta}}$ \cite{Acin_2016, Woodhead_2020}, realizing what is sometimes called ``post-hoc" self-testing \cite{Supic_2020}. 
In case of extremal POVMs, this has be proven to be enough to compute the guessing probability of the eavesdropper is given by \cite{Brown_2020}
\begin{equation}
    p_g=\max_{\{\ket{\psi},\Pi_b,E_b\}}\sum_b \expval{\openone \otimes \Pi_b\otimes E_b}{\psi}\,.
\end{equation}
where the maximum is evaluated over all possible $\ket\psi$ and $\Pi_b$ that satisfy Eq.\ \eqref{eq:constr_mean_value}.
The guessing probability bounds the number of secure bits through the min-entropy, defined as $H_{min} = -\log_2 p_g$.
However, to compute other quantities, such as the worst-case conditional von Neumann entropy, it is crucial to establish conditions on how $\Pi_b$ acts on the state $\ket{\psi}$, rather than relying solely on the constraints \eqref{eq:constr_mean_value} imposed on the mean value. 
This can be achieved by introducing a saturated Tsirelson bound, involving the operators $\Pi_b$.

\subsection{Tsirelson bounds for rank-one POVMs}
\label{sec:TsirelboundsforPOVM}
Our first result is the identification of
a Tsirelson bound, i.e., an upper limit to quantum mechanical correlations, that is saturated when Bob performs the measurement by using rank-one POVM $\{F_b\}$.
A key observation is that, for a couple of generic projectors $\Lambda_b'$ and $\Pi_b'$ on Alice and Bob's sides, respectively, it holds:
\begin{equation}
\label{eq:ineq_bound}
    \expval{\Lambda_b' \otimes \Pi_b'}=
\expval{(\Lambda_b' \otimes \Pi_b')^2}=
\|\Lambda_b' \otimes \Pi_b'\ket{\psi}\|^2\geq 0 \ .
\end{equation}
Therefore if, relying on the self-testing properties \eqref{eq:op_alice}, we find a combination $\Lambda_b$ of Alice's operators satisfying
\begin{align}
\label{eq:piece_bound}  
    \expval{\Lambda_b \otimes \Pi_b}{\psi}&=0\,,\\
     \Lambda_b^2\ket{\psi}&=\Lambda_b\ket{\psi}\qquad
    \forall b\,, \label{eq:piece_bound2}  
\end{align}
we can conclude that the correlations $\expval{\Lambda_b \otimes\Pi_b}$ lie on a boundary of the quantum set, identified by \eqref{eq:piece_bound} and by the tilted CHSH inequalities used to derive \eqref{eq:op_alice}. 
Note that the second term in \eqref{eq:ineq_bound} can be also interpreted as a sum of squares (SOS) decomposition made of a single term, and the saturation of the bound implies 
\begin{equation}
\label{eq:state_cond}
    \Lambda_b \otimes \Pi_b\ket{\psi} = 0\,.
\end{equation}

If we consider a distribution generated by a rank-one POVM $\{F_b\}$ acting on $\ket{\phi_{\theta}}$ (see Eq.\ \eqref{eq:constr_mean_value}), the explicit form of $\Lambda_b$ can be derived, as described in Appendix \ref{appendix:coefficients}. 
By parameterizing the vectors $\ket{\beta_b}$ in \eqref{eq:F_proj} in terms of the Bloch angles $\theta_b^B$ and $\phi_b^B$
\begin{equation}
    \label{eq:beta}
    \ket{\beta_b}=\cos(\frac{\theta^B_b}{2})\ket0+
e^{i\phi^B_b}\sin(\frac{\theta^B_b}{2})\ket1 \,,
\end{equation}
the explicit expression of $\Lambda_b$ is given by
\begin{equation}
\label{eq:comb_A}
    \Lambda_b= \frac12(\openone-\vec{n}_{b}\cdot \vec{A})\,,\qquad |\vec{n}_b|^2=1
\end{equation}
where $\vec{A}=(A_1, A_2,A_3)$ and
\begin{equation}
\label{eq:final_coeff}
\begin{split}
    n_{1b}&=\frac{\sin \theta \sin \theta^B_b}{1+\cos \theta \cos \theta^B_b}\cos\phi^B_b  \ , \\
    n_{2b}&=-\frac{\sin \theta \sin \theta^B_b}{1+\cos \theta \cos \theta^B_b}\sin\phi^B_b \ , \\
    n_{3b}&=\frac{\cos \theta+\cos \theta^B_b}{1+\cos \theta \cos \theta^B_b} \ .
\end{split}
\end{equation}

Thanks to the self-testing \eqref{eq:op_alice}
and the condition $|\vec{n}_{b}|^2=1$, $\Lambda_b$ is also guaranteed to satisfy \eqref{eq:piece_bound2}. 
We note that, in deriving the expression for $n_{2b}$ we assumed the case where Alice implements $+\sigma_2$, which explains the minus sign preceding the expression.
However, we remark that, in a DI scenario, the indeterminacy of $\sigma_2$, explained after \eqref{eq:op_alice}, remains because we do not know the orientation of $\sigma_2$ on Bob's side.

With this parametrization, the condition \eqref{eq:state_cond} becomes
\begin{equation}
\label{eq:Tsirelson_saturated}
   (\vec{n}_{b}\cdot \vec A) \otimes \Pi_b\ket{\psi}=\Pi_b\ket{\psi}\,,\quad\forall b \ .
\end{equation} 
We can express this set of boundary equations in a compact form by summing over $b$ and introducing a Bell operator $\mathcal{S}$,
\begin{equation}
\label{eq:T_op}
    \mathcal S \equiv\sum_b (\vec{n}_{b}\cdot \vec A) \otimes \Pi_b\,,
\end{equation}
satisfying the boundary condition
\begin{equation}
\label{eq:T_cond}
    \mathcal{S}\ket{\psi} = \ket{\psi} \iff \expval{\mathcal{S}}=1\,.
\end{equation}
The latter is equivalent to \eqref{eq:Tsirelson_saturated}, since, by multiplying by $\Pi_{b}$, we obtain the equations \eqref{eq:Tsirelson_saturated}.
Note that the mean value $\expval{\mathcal{S}}$ represents a hyperplane in the space of quantum correlations, for which, from Eq.\ \eqref{eq:ineq_bound}, we find the Tsirelson bound:
\begin{equation}
\label{eq:T_ineq}
    \expval{\mathcal{S}} \leq 1\,.
\end{equation}

We informally summarize our first result as follows:
\begin{result}
    Any rank-one qubit POVM measured on any two-qubit entangled state can generate correlations, with spatially separated self-tested measurements, that lie on the border $\expval{\mathcal S}=1$ of the set of quantum correlations.
\end{result}
We remark that this result holds regardless of whether the POVM is extremal. 
The distinction between extremal and non-extremal POVMs instead lies in the extremality of the generated correlations and plays a crucial role in determining the amount of randomness that can be extracted from the measurements \cite{Franz2011}. 

\section{Randomness estimation}
\label{sec:Randomness_Estimation}
In the previous section, we associated a boundary, given by Eq.\ \eqref{eq:piece_bound}, to every rank-one POVM through the parameters $\vec{n}_b$. 
This association proves to be both useful and insightful in the process of randomness estimation, as it highlights the differences between extremal and non-extremal POVMs.

We rewrite the Tsirelson bound \eqref{eq:T_cond} in terms of the  identity and of the operators $M_j$ ($j=1,2,3$): 
\begin{equation}
\label{eq:PitoB}
\openone = \sum_{b=0}^{d-1} \Pi_b \ ,  \qquad M_j =\sum_{b=0}^{d-1} n_{jb}\Pi_b \ ,  
\end{equation}
so that 
\begin{equation}
\label{eq:bound_condition_state}
 \sum_{j=1}^3 A_j \otimes M_j\ket{\psi}=\ket{\psi} \ .
\end{equation}
Taking the scalar product of this expression with $A_i\otimes\openone \otimes E_e \ket{\psi}$, for a generic hermitian operator $E_e$ of Eve, we get $\sum_{j=1}^3 \expval{(A_i A_j) \otimes M_j \otimes E_e}{\psi}= \expval{A_i\otimes \openone \otimes E_e}{\psi}$
and, expanding the sum,
\begin{equation}
\label{eq:mean_values_expand}
    \expval{M_i\otimes E_e}+\sum_{j \neq i}\expval{(A_iA_j)\otimes M_j \otimes E_e}=
\expval{A_i \otimes E_e} \ .
\end{equation}
Looking at \eqref{eq:op_alice} and recalling that operators of different parties commute, we observe that the
two terms $\expval{M_i\otimes E_e}$ and $\expval{A_i \otimes E_e}$ in \eqref{eq:mean_values_expand} are real (being mean values of Hermitian operators) while the others are imaginary (mean values of anti-Hermitian operators). 
So we can conclude that 
\begin{equation}
\label{eq:sep_cond}
    \expval{M_i \otimes E_e} =  \expval{A_i\otimes E_e}\,. 
\end{equation}
In particular, by choosing $E_e=\openone_E$, the above result implies that 
$\expval{M_i}=\expval{A_i}$.
Using this observation and the factorization \eqref{eq:fact_alice_eve}, we find
\begin{equation}
\label{eq:fact_B_E}
    \expval{M_i \otimes E_e}=\expval{A_i \otimes E_e}=\expval{A_i}\expval{E_e}=\expval{M_i}\expval{E_e} \,.
\end{equation}
In summary, given a generic rank-one POVM, we derived a Tsirelson bound involving the correlations $\expval{A_j\otimes \Pi_b}$. When the bound is saturated,  we found a combination $M_j$ of the operators $\Pi_b$ (see Eq.\ \eqref{eq:PitoB}) such that the joint probability distribution of $M_j$  with a generic measurement of Eve $E_e$ factorizes, as shown in Eq.\ \eqref{eq:fact_B_E}.
This equation is the key relation upon which our second main result relies: In the next section, we show how this factorization allows us to evaluate the randomness generated  by the $\{\Pi_b\}$ outcomes.

\subsection{Randomness for extremal POVMs}
\label{sec:randomness_extremal_povm}
\begin{figure}
    \centering
    \includegraphics[width=.65\linewidth]{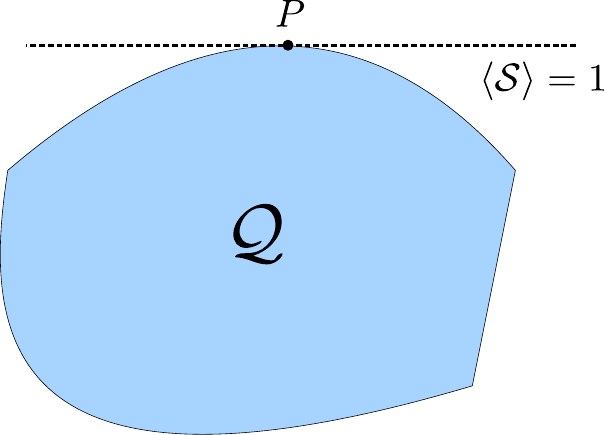}
    \caption{Qualitative representation of the non-local portion of quantum correlations identified by the self-testing \eqref{eq:op_alice}. When the POVM is extremal, the saturation of the Tsirelson bound $\expval{\mathcal S}\leq 1$ identifies an extremal point $P$ on the boundary.}
    \label{fig:Q_extremal}
\end{figure}

Extremality of the POVM generating the correlations is equivalent to the invertibility of equations \eqref{eq:PitoB} (see Appendix \ref{appendix:lin_ind}), namely to the possibility of rewriting every $\Pi_b$ in terms of $\openone$ and $M_j$. Then, if the relations in \eqref{eq:PitoB} are invertible, 
we can conclude that, for each outcome $b$ 
\begin{equation}
\label{eq:fact_Pi_E}
    \expval{\Pi_b E_e}=\expval{\Pi_b}\expval{E_e}\,
\end{equation}
and all Bob's and Eve's outcomes are uncorrelated.
We note that, from \eqref{eq:fact_Pi_E}, when extremal POVMs are considered, the saturation $\expval{\mathcal{S}}=1$ implies that the correlations between Alice's measurements and Bob's POVM are extremal in the quantum set, meaning they cannot be reproduced as a convex sum of other correlations in the set \cite{Franz2011} (see Fig.\ \ref{fig:Q_extremal}). 
Indeed, the saturation allows us to deduce all the correlations between Alice and Bob: By replacing $E_e = \openone_E$ and leveraging the invertibility discussed earlier, we can compute the marginals $\expval{\Pi_b}$ from \eqref{eq:sep_cond} and the correlations with Alice through \eqref{eq:Tsirelson_saturated}. 
This proves that the correlations satisfying $\expval{\mathcal{S}}=1$ are unique and therefore extremal.

For correlations satisfying \eqref{eq:fact_Pi_E}, the guessing probability can be simply computed as
\begin{equation}
p_g=\max_b\expval{\Pi_b}\,.  
\end{equation}
Indeed
\begin{equation}
\begin{aligned}
\label{eq:classicalguess}
    p_g&=\max_{\{E_b\}}\sum_b \expval{\Pi_b\otimes E_b}=\max_{\{E_b\}}\sum_b \expval{\Pi_b}\expval{E_b}
    \\
    &\leq\max_b\expval{\Pi_b}\max_{\{E_b\}}\sum_b \expval{E_b} = \max_b\expval{\Pi_b} \ .
\end{aligned}    
\end{equation}
However, we highlight that the randomness is derived, in a device independent scenario, just imposing the saturation of some Bell inequality necessary to deduce \eqref{eq:op_alice}, and the boundary condition \eqref{eq:bound_condition_state}, without the need of explicitly verify the entire joint probability distribution.

Also, starting from the factorization \eqref{eq:fact_Pi_E}, we can compute the conditional von Neumann entropy to quantify the randomness. 
Indeed, after Bob realizes his measurement, he holds a classical register made of all the possible outcomes weighted by their probability to happen. 
Eve, however, could hold some side information on this result, encoded in the state $\rho_{E}^b$ which is left after the action of $\Pi_b$ on the initial state (and tracing away Alice). 
This can be summarized by introducing the classical-quantum post measurement state, defined as
\begin{equation}
\label{eq:post_meas}
    \rho^{post}_{B E}=\sum_{b} p_B(b)\ketbra{b}{b}\otimes \rho_E^{b}
\end{equation}
with $p_B(b) \equiv \expval{\Pi_b}$ and
\begin{equation}
    \rho_E^{b}=\frac{\Tr_{AB}\Bigl[(\openone \otimes \Pi_b \otimes \openone) \ketbra{\psi}\Bigr]}{p_B(b)} \,,
\end{equation} 
a normalized state on Eve's space.
The post-measurement state \eqref{eq:post_meas} captures potential correlations between Eve and the classical register containing Bob's outcomes. 
To quantify the extractable randomness, one can evaluate the conditional von Neumann entropy on this state.

In practice, Eve may gain information from the outcome $e$ of a general measurement $E_e$, with the associated marginal probability defined as $p_E(e) \equiv \expval{E_e}$ and the correlation with Bob as $p_{BE}(e,b) \equiv \expval{\Pi_b E_e}$. 
However, if the observed distribution by Bob and Alice arises from an extremal POVM, the factorization \eqref{eq:fact_Pi_E} can be applied, allowing us to write
\begin{equation}
\label{eq:prob_cond_noncond}
\begin{split}    
    \Tr_E\Bigl[E_e \rho_E^b\Bigr]&=p_E(e\lvert b) =
    \frac{p_{BE}(e,b)}{p_B(b)} \\
    &=p_E(e)=\Tr_E\Bigl[E_e \Tr_{AB}\Bigl[\ketbra{\psi}\Bigr]\Bigr] \
\end{split}
\end{equation}
which is telling us that the result of Eve measurement does not depend on Bob's results.
Since the previous expression holds for every possible choice of $E_e$, it follows that $\rho^b_E=\Tr_{AB}\Bigl[\ketbra{\psi}\Bigr]$ and the post-measurement state \eqref{eq:post_meas} factorizes into separate parts for Bob and Eve:
\begin{equation}
\label{eq:post_meas_fact}
    \rho^{post}_{BE}=\left(\sum_{b}p_B(b)\ketbra{b}{b}\right)\otimes \Tr_{AB}\Bigl[\ketbra{\psi}\Bigr]\equiv \rho_B^{post}\otimes \rho_E^{post}.
\end{equation}
It is then immediate to conclude about the conditional von Neumann entropy. 
Indeed, using the additive property
\begin{equation}
    H(B|E)_{\rho^{post}_{BE}} = H(BE)_{\rho^{post}_{BE}} - H(E)_{\rho^{post}_{BE}} = H(B)_{\rho^{post}_{BE}}\,.
\end{equation}
From the diagonal form of $\rho_B^{post}$ it is then easy to identify $H(B)_{\rho^{post}_{BE}}$ with the Shannon entropy of the probability distribution of the POVM outcomes.

We informally summarize our second result as follows:
\begin{result}
    In our scenario, when correlations generated by an extremal qubit POVM are considered, the saturation $\expval{\mathcal{S}}=1$ implies the uniqueness of the correlations and their extremality on the border of the quantum set.
    In this ideal case, the device-independent randomness coincides with that of a trusted scenario.
\end{result}
We anticipate that the mean value $\expval{\mathcal{S}}$ is useful also when noisy correlations are observed.
We will discuss this case in a specific example in Sec.\ \ref{sec:3_out}.

\subsection{Randomness for non-extremal POVMs}
If the correlations are generated by a non-extremal POVM, relations \eqref{eq:PitoB} are non-invertible. 
A simple example arises when Bob's measurements have more than four outcomes (i.e., $b = 0, \dots, d-1$ with $d > 4$): In this case, the POVM is non-extremal \cite{D'Ariano_2005}, and there cannot be a one-to-one correspondence between $\{\Pi_b\}$ and $\{ \openone,M_j \}$. 
Consequently, the separability of the correlations \eqref{eq:fact_B_E} between $M_j$ and $E_e$ does not imply the separability of \eqref{eq:fact_Pi_E} between $\Pi_b$ and $E_e$. 
As a result, the post-measurement state is not separable, and even the simple argument used in \eqref{eq:classicalguess} to compute the guessing probability cannot be applied.

Geometrically, the situation is as follows: Eq.\ \eqref{eq:T_cond} defines a boundary where the correlations generated by the non-extremal POVM $F_b$ reside. Unlike in the extremal case, the intersection between the hyperplane described by \eqref{eq:T_cond} and the quantum set is not a single point in general. 
Indeed, being $F_b$ non-extremal,
\begin{equation}
\begin{split}
    \expval{\Lambda_b \otimes F_b} & =\sum_k p_k \expval*{\Lambda_b \otimes F_b^{(k)}}=0
\end{split}
\end{equation}
for some extremal POVMs $\{ F_b^{(k)} \}$ (which are necessarily rank-one) and some probabilities $p_k >0$. The previous equations can only be satisfied if all terms in the sum are zero, since each term is positive.
Consequently, the same boundary conditions \eqref{eq:T_cond} apply to the correlations generated by each $F_b^{(k)}$, with each representing an extremal point in the quantum set. 
Therefore, while for extremal POVMs the boundary conditions fully determine the set of correlations, this does not hold for non-extremal POVMs.
In terms of randomness, this means that the randomness obtainable from the complete set of correlations differs from that which can be enforced solely through the boundary conditions \eqref{eq:T_cond}. 
By imposing only the latter constraint, Eve retains the freedom to choose the most advantageous combination of extremal correlations points on the boundary.
Nevertheless, the boundary condition can sometimes suffice to establish a lower bound on the extractable randomness. 
A concrete example of this approach is provided in Sec.\ \ref{sec:non_ext_4}.

Finally, we note that in certain realizations, additional physical constraints may arise, as in the sequential scenario discussed in Sec.\ \ref{sec:sequential}. 
These extra constraints can alter the geometry of the quantum set, thereby affecting the number of intersection points with the hyperplane \eqref{eq:T_cond}. 
This, in turn, has direct implications for the extremality of the correlations and the amount of extractable randomness.

\section{Examples}
\label{sec:examples}
\subsection{Extremal 3-outcome POVM}
\label{sec:3_out}
\begin{figure}
    \centering
    \includegraphics[width=\linewidth]{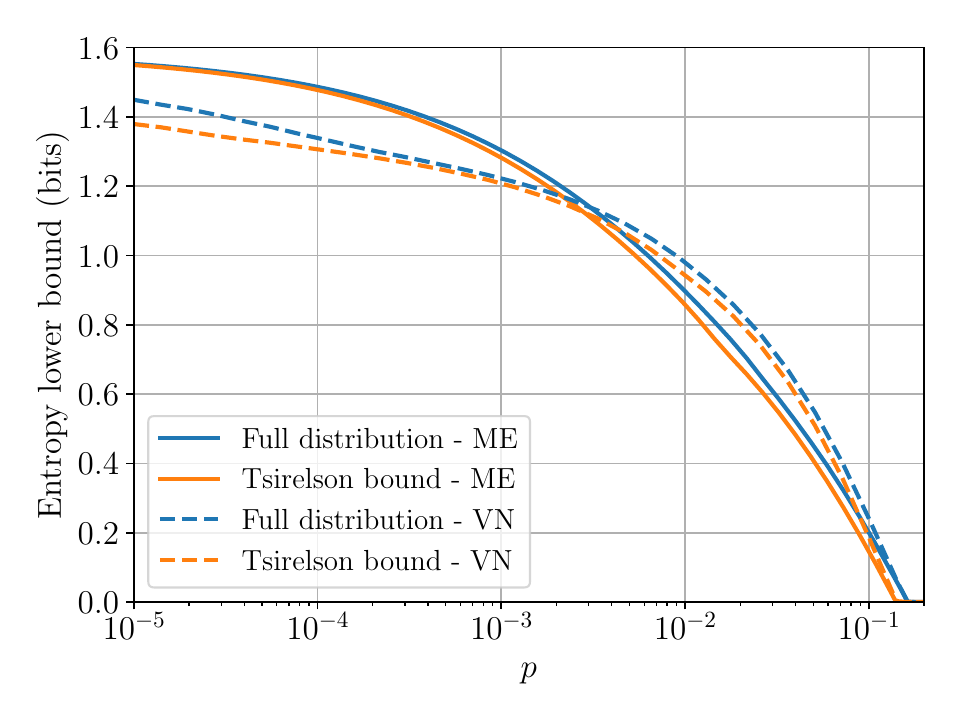}
    \caption{Lower bounds on the min-entropy (ME) and von Neumann entropy (VN) for the cases where, given a noise term $p$ and a CHSH value of $2\sqrt{2}(1-p)$, the full distribution of the extremal triangular POVM is constrained and the case where only the Tsirelson bound \eqref{eq:baptista_bound}, equal to $1-p$, is imposed.   
    The parameter $p$ is assumed to be a white noise contribution on the ideal maximally entangled state ($\theta = \frac{\pi}{2}$); see also Section \ref{sec:exp_results}.
    The computation of the min-entropy bound is performed with the numerical methods described in Appendix \ref{sec:randomness_estimation_app}, with an NPA order 3 and the solver SDPA-DD \cite{nakata2010numerical}, while the computation of the von Neumann entropy bound with an NPA order 2 and solver SDPA.  The average rate loss when only the Tsirelson bound is considered is approximately 7\% for both entropies.
    For a given value of $p$, the higher lower bound should be chosen.}
    \label{fig:scanp}
\end{figure}

\begin{figure*}
    \centering
    \includegraphics[width=0.9\textwidth]{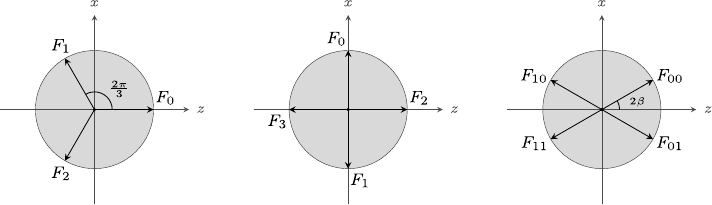}
    \caption{The three POVMs, represented on the Bloch plane, considered in the examples of Sections \ref{sec:3_out}, \ref{sec:non_ext_4} and \ref{sec:sequential}, respectively.}
    \label{fig:povmexamples}
\end{figure*}
As first simple case, we consider the situation in which the correlations are produced by the action on the state \eqref{eq:state_Bloch} of the extremal POVM
\begin{equation}
\begin{aligned}
\label{eq:mercedesPOVM}
    F_0 &=\frac{2}{3}\ketbra{0} \ , \\
    F_1 &=\frac{2}{3} R\qty(\frac{\pi}{3})\ketbra{0}R\qty(-\frac{\pi}{3}) \ , \\
    F_2 &=\frac{2}{3} R\qty(-\frac{\pi}{3})\ketbra{0}R\qty(\frac{\pi}{3}) \,,
\end{aligned}
\end{equation}
where we introduced the rotation matrix
\begin{equation} 
    R(\alpha)=\mqty(\cos \alpha & -\sin \alpha \\ \sin \alpha & \cos \alpha) \,.
\end{equation}
See Fig.\ \ref{fig:povmexamples} for a Bloch sphere representation of this POVM.
In this case, the relation \eqref{eq:PitoB} becomes
\begin{equation}
\label{eq:rel_3out}
    \mqty(\openone \\ M_1 \\ M_3)= \mqty(1 & 1 & 1 \\
    0 &  \frac{\sqrt{3}\sin \theta}{2-\cos \theta }& -\frac{\sqrt{3}\sin \theta}{2-\cos \theta } \\
    1 & \frac{1-2 \cos \theta}{\cos \theta -2} & \frac{1-2 \cos \theta}{\cos \theta -2} ) \mqty(\Pi_0 \\ \Pi_1 \\ \Pi_2) \ .
\end{equation}
Note that, as expected, the matrix in \eqref{eq:rel_3out} can be inverted. Then, from the discussion in Sec.\ \ref{sec:randomness_extremal_povm}, we can immediately conclude that the post measurement state \eqref{eq:post_meas} is separable. Therefore, the conditional von Neumann entropy matches the Shannon entropy of the outcomes and the guessing probability matches the probability of the most probable outcome, as in \eqref{eq:classicalguess}.

The Tsirelson bound can be derived by applying \eqref{eq:bound_condition_state}:
\begin{equation}
\label{eq:mercedes_bound}
\begin{split}
    1 = & \frac{\sqrt{3}\sin{\theta}}{2-\cos{\theta}}\expval{A_1\otimes(\Pi_1-\Pi_2)}
    + \expval{A_3 \otimes \Pi_0} \\
    &+ \frac{1-2\cos{\theta}}{\cos{\theta}-2}\expval{A_3 \otimes (\Pi_1+\Pi_2)} \,.
\end{split}
\end{equation}
Note also that, in the special case $\theta=\pi/2$, this bounds becomes
\begin{equation}
\label{eq:baptista_bound}
   \frac{\sqrt{3}}{2} \! \expval{A_1\otimes(\Pi_1 \!-\Pi_2)} +  \expval{A_3 \otimes \Pi_0} - \frac{1}{2} \!  \expval{A_3\otimes(\Pi_1\! +\Pi_2) } \! = \! 1
\end{equation}
in agreement with the results in Sec.\ 6 of \cite{baptista2023}. 

For this POVM, we performed numerical simulations to lower bound the certifiable min-entropy and von Neumann entropy when some random noise $p$ affects the ideal quantum state (for more details, see the discussion around Eq.\ \eqref{eq:exp_state}, setting $\theta=\frac{\pi}{2}$ and $c=0$). 
Our analysis compares the bounds on the two entropies obtained when all correlations are considered with the case in which only the average value of $\mathcal{S}$ is constrained. 
In the simulations, given an average CHSH operator value of $2\sqrt{2}(1-p)$, we compared the outcomes obtained by either constraining solely the average value of the operator $\mathcal{S}$ (equal to $1-p$) or enforcing the full correlations between Alice and the POVM. 
The results, shown in Fig.\ \ref{fig:scanp}, reveal an average loss of min-entropy of 7\% when only $\expval{\mathcal{S}}$ is constrained for both entropies.
The min-entropy computations are carried out using the NPA hierarchy \cite{navascues_bounding_2007, navascues_convergent_2008} at order 3 and the SDPA-DD solver with a precision of $10^{-12}$ \cite{nakata2010numerical}\, while the von Neumann computations are carried out using the NPA hierarchy at order 2 and the SDPA solver with a precision of $10^{-7}$ (see also Appendix \ref{sec:randomness_estimation_app}).
We emphasize that, in general, the true von Neumann entropy is always greater than the min-entropy.
The fact that the lower bounds in Fig.\ \ref{fig:scanp} cross each other may arise from the different ways in which the two entropies are bounded.
Although both methods rely on the NPA hierarchy, they are inherently distinct.
Furthermore, since computing the von Neumann entropy bound is more demanding in terms of computational resources, we have chosen a lower order of the NPA hierarchy and a less precise numerical solver than min-entropy computations; moreover we adopted some strategies from \cite{Brown2024} to achieve faster (though less tight) computations.
In a practical QRNG implementation, these considerations should be taken into account when estimating the final rate, and the method estimating the higher lower bound should be preferred.
To verify the feasibility in a real word scenario of a protocol involving this 3-outcome POVM,  we have conducted a proof-of-concept quantum optics experiment (see Sec.\ \ref{section:experimental_implementation}).

\subsection{Non-extremal 4-outcome POVM}
\label{sec:non_ext_4}
As previously mentioned, when correlations are generated by a non-extremal POVM, the results on the separability of the post-measurement state do not apply. 
However, we will demonstrate through a simple yet meaningful example that, in some cases, conclusions about the guessing probability can still be drawn by leveraging the boundary condition \eqref{eq:bound_condition_state}, even in non-extremal scenarios.
We consider the 4-outcome POVM (see Fig.\ \ref{fig:povmexamples} for a Bloch sphere representation) with elements
\begin{align}
\label{eq:POVM_4out_non}
    F_0&=\frac{1}{2}\ketbra{+}{+} \ , && & F_1&=\frac{1}{2}\ketbra{-}{-} \ , \\
    F_2&=\frac{1}{2}\ketbra{0}{0} \ ,  && &F_3&=\frac{1}{2}\ketbra{1}{1} 
\end{align}
acting on the entangled state $\ket{\phi_{\theta}}$.
The above POVM is non-extremal, as it can be directly verified with the methods introduced in \cite{D'Ariano_2005} or by noting that it can be (uniquely) decomposed in terms of extremal POVMs as
\begin{equation}
\label{eq:decompQuadrato}
    \mqty(F_0 \\ F_1 \\ F_2 \\ F_3) = \frac{1}{2}\mqty(2F_0\\ 2F_1 \\ 0 \\0) + \frac{1}{2}\mqty( 0 \\0 \\ 2F_2\\ 2F_3) \ .
\end{equation}
Indeed, if we try to perform our DI procedure and introduce the relations, as in \eqref{eq:PitoB},
\begin{equation}
\label{eq:rel_4out}
    \mqty(\openone \\ M_1 \\ M_2 \\ M_3)= \mqty(1 & 1 & 1 & 1 \\ 
          \sin \theta & -\sin \theta & 0 & 0 \\ 0 & 0 & 0 & 0 \\ \cos \theta & \cos \theta & 1 & -1 ) \mqty(\Pi_0 \\ \Pi_1 \\ \Pi_2 \\ \Pi_3)
\end{equation}
we can still conclude that the correlations of $M_j$ and Eve are separated, but then we cannot invert the relations between $M_j$ and $\Pi_b$ in \eqref{eq:rel_4out}. 
Hence, the entire discussion done for the entropies in Sec.\ \ref{sec:randomness_extremal_povm} does not apply. 

To bypass this problem, we can draw inspiration from the decomposition in \eqref{eq:decompQuadrato} and divide the boundary
conditions given in \eqref{eq:Tsirelson_saturated}
in two equations:
\begin{align}
    &\sum_j \left(n_{j0} A_j \Pi_0+ n_{j1}A_j \Pi_1 \right)\ket{\psi}=(\Pi_0+\Pi_1)\ket{\psi} \ , \\
   &\sum_j \left( n_{j2} A_j \Pi_2+n_{j3}A_j \Pi_3 \right)\ket{\psi}=(\Pi_2+\Pi_3)\ket{\psi} 
\end{align}
which are ideally associated with the extremal POVMs in the decomposition \eqref{eq:decompQuadrato}. The two operators appearing on the right hand sides
\begin{equation}
    M_0^{(1)} =\Pi_0+\Pi_1 \ , \quad  M_0^{(2)} =\Pi_2+\Pi_3 \
\end{equation}
describe the subspaces on which the elements of the decomposition live or, from another perspective, they are associated with two different strategies that an eavesdropper can combine to reproduce the observed correlations. We then expect to bound the guessing probability by writing it in terms of $M_j$ (whose correlations are separated by Eve) and  $M_0^{(1)}$, $M_0^{(2)}$ (which are related to the freedom arising from the non-extremality of the POVM). 
For this reason, we first note that
\begin{align}
    \Pi_0& =\frac{M_0^{(1)}\sin \theta +M_1}{2 \sin \theta} \ , \\ 
    \Pi_1&= \frac{M_0^{(1)}\sin \theta -M_1}{2 \sin \theta}  \ , \\
    \Pi_2& =\frac{M_0^{(2)}+M_3-\cos\theta M_0^{(1)}}{2} \ ,  \\
    \Pi_3&= \frac{M_0^{(2)}-M_3+\cos \theta M_0^{(1)}}{2} 
\end{align}
and we insert these relations in the definition of the guessing probability.
Then, with some manipulations described in Appendix \ref{sec:app_eve_strat}, one can actually prove that the guessing probability satisfies
\begin{equation}
\label{eq:guess_non_extr}
    p_g = \max_{\text{E}}\sum_b \expval{\Pi_b E_b} \le \frac{1}{2}+\frac{\abs{\cos \theta}}{2} \ .
\end{equation}
By explicitly constructing a strategy for the eavesdropper, as detailed in Appendix \ref{sec:app_eve_strat}, it can be verified that the previous inequality can be saturated.

The guessing probability matches what one would find in a 2-outcome scenario where the correlations are realized using the right component of the decomposition in \eqref{eq:decompQuadrato}. 
Indeed, the saturating strategy corresponds to a scenario in which Eve prepares the devices such that only the right component of the decomposition is implemented. 
While this strategy does not reproduce the entire set of correlations obtainable in the experiment, it satisfies the boundary equation \eqref{eq:T_cond}, which is the only constraint imposed.

\begin{figure}
    \centering
    \includegraphics[width=.65\linewidth]{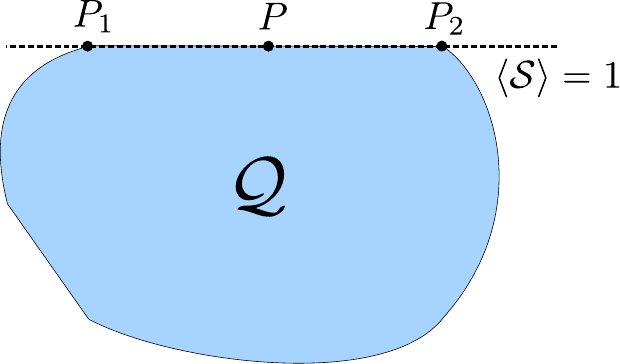}
    \caption{Qualitative representation of the non-local portion of quantum correlations identified by the self-testing \eqref{eq:op_alice} when the non-extremal 4-outcome POVM is considered. In this case, saturating the bound is not associated with a single point in the quantum correlations set; rather, at least two possible extremal points are the correlations generated by the two extremal POVMs on the RHS of the decomposition in \eqref{eq:decompQuadrato}.}
    \label{fig:Q_squarePOVM}
\end{figure}
To gain a geometrical intuition of the problem, refer to Fig.\ \ref{fig:Q_squarePOVM}, where the situation is sketched. 
Every combination of the correlations $P_1$ and $P_2$ generated by the two extremal POVMs in \eqref{eq:decompQuadrato} satisfies the same boundary equation, represented by the dashed line. 
The POVM in \eqref{eq:POVM_4out_non} corresponds to the combination $P$, where the two correlations $P_1$ and $P_2$ are equally weighted. 
Without additional constraints, such as $\expval{\Pi_0 + \Pi_1} = \expval{\Pi_2 + \Pi_3}$, the boundary condition allows Eve to select the most favorable strategy along the boundary, which turns out to be $P_2$. 
For further details, refer to Appendix \ref{sec:app_eve_strat}.

\subsection{Sequential POVM}
\label{sec:sequential}

\begin{figure}
    \centering
    \includegraphics[width=.65\linewidth]{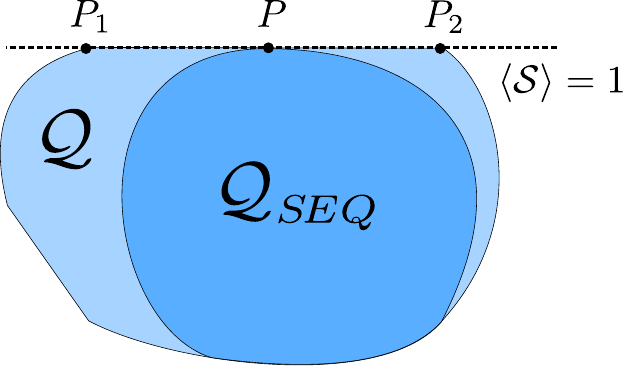}
    \caption{Qualitative representation of the non-local portion of quantum correlations identified by the self-testing \eqref{eq:op_alice} when the non-extremal sequential POVM is considered. As discussed in the main text, this POVM does not produce extremal correlations in the set of quantum correlations but it does in the set of sequential quantum correlations.}
    \label{fig:Q_Qseq}
\end{figure}

A rank-one POVM could also arise from sequential measurements, as in the scenario discussed in \cite{Padovan:2023mxc}. 
In that case, a first user, call it Bob$_1$, performs a measurement described by the POVM 
\begin{equation}
\begin{split}
   F_{0}^1&=K_0^\dagger K_0 \ , \qquad K_0=\cos \beta \ketbra{0}+\sin \beta \ketbra{1} \ , \\ 
    F_{1}^1&=K_1^\dagger K_1 \ , \qquad K_1=\cos \beta \ketbra{1}+\sin\beta \ketbra{0}\,,
\end{split}
\end{equation}
on the state $\ket{\phi^+}$. 
The resulting state is then passed to a second user, Bob$_2$, who applies
\begin{equation}
\begin{split}
    F_0^2&=\ketbra{+} \ , \\
    F_1^2&=\ketbra{-} \ .
\end{split}
\end{equation}
This sequential action can be rewritten in terms of a unique 4-outcome POVM on the initial state $\ket{\phi^+}$, with elements $F_{b_1,b_2} \equiv K_{b_1}^\dagger F^2_{b_2} K_{b_1}$:
\begin{equation}
\label{eq:seq_POVM}
    F_{b_1,b_2}=\frac{1}{2}\left(\frac{\openone+(-1)^{b_2}\sin (2 \beta) \sigma_1+(-1)^{b_1}\cos (2 \beta) \sigma_z}{2} \right)
\end{equation}
where $b_1, b_2=0,1$ label the element of the POVM of Bob$_1$ and Bob$_2$, respectively. 
See Fig.\ \ref{fig:povmexamples} for a Bloch sphere representation of the 4-outcome POVM.
This POVM is made by rank-one elements, and therefore we can apply the discussion of Sec.\ \ref{sec:TsirelboundsforPOVM}, obtaining the condition
\begin{equation}
\begin{split}
    \ket{\psi} =\,& \sin\!\left(2\beta\right)\, A_1 \big(\Pi_{0,0}+\Pi_{1,0}-\Pi_{0,1}-\Pi_{1,1}\big)\ket{\psi}\\
    &+ \cos\!\left(2\beta\right)\, A_3 \big(\Pi_{0,0}+\Pi_{0,1}-\Pi_{1,0}-\Pi_{1,1}\big)\ket{\psi}\,.\end{split}
\end{equation}
which, up to a difference in notation, is the same condition found in \cite{Padovan:2023mxc} (see Proposition 3 in the supplementary material therein). 

Note that the POVM in \eqref{eq:seq_POVM} is non-extremal, indeed it can be decomposed in a very similar way to the one of the previous example:
\begin{equation}
\label{eq:decompSeq}
    \mqty(F_{0,0} \\ F_{0,1} \\ F_{1,0} \\ F_{1,1}) = \frac{1}{2}\mqty(2F_{0,0}\\ 0\\ 0 \\2F_{1,1} ) + \frac{1}{2}\mqty( 0 \\2F_{0,1}  \\ 2F_{1,0} \\ 0)\,,
\end{equation}
and therefore we cannot certify the maximal randomness, which is 2 bits, in general. 
However, this maximal certification is possible (along with the extremality of the correlations \cite{Padovan2026}) if we consider additional physical constraints stemming from the sequential scenario \cite{Padovan:2023mxc}.

In Fig.\ \ref{fig:Q_Qseq}, we sketch an intuitive geometrical interpretation of this discussion: a portion of the quantum set $\mathcal{Q}$ is depicted.
The point $P$ represents the correlations of the sequential POVM on the LHS in \eqref{eq:decompQuadrato}, while $P_1$ and $P_2$ are the correlations when the first or the second term on RHS in \eqref{eq:decompQuadrato} are implemented, respectively.
As previously stated, the correlations of the sequential POVM are not extremal in $\mathcal{Q}$, but are extremal in the set of sequential quantum correlations $\mathcal{Q}_{SEQ}$.

\section{Experimental implementation of 3-outcome POVM}
\label{section:experimental_implementation}
The aim of our proof-of-concept experiment is to show that the correlations obtainable from the real-world implementation of the state and measurements introduced in Sec.\ \ref{sec:3_out} could be used for device-independent randomness applications.
The idea is to exploit the violation of a tilted CHSH inequality and the mean value of $\mathcal{S}$ in \eqref{eq:mercedes_bound}.
Note that we work under fair-sampling assumption without addressing the presence of loopholes and without actively choosing the inputs of Alice and Bob. 

\subsection{Protocol}
\label{subsection:protocol}
In the ideal protocol, Alice and Bob share the partially entangled state $\ket{\phi_{\theta}}$ given in Eq.\ \eqref{eq:state_Bloch}. 
Alice’s ideal measurements are associated with the dichotomic observables
\begin{equation}
    \aliceobs{1} = \sigma_1 \ , \qquad \aliceobs{3} = \sigma_3\,
\end{equation}
and Bob's ideal measurements are
\begin{align}
    \bobobs{0} &= \cos\mu \sigma_z + \sin\mu \sigma_1 \ , \\ 
    \bobobs{1} &= \cos\mu \sigma_z - \sin\mu \sigma_1 \ , \\ 
    \mathcal{B}_{2} &= \qty{F_b}_{b=0,1,2} \,,
\end{align}
where the set $\qty{F_b}_b$ represents the triangular POVM introduced in \eqref{eq:mercedesPOVM}, with outcomes labeled by $b=0,1,2$, and where the parameter $\mu$ is defined as $\mu = \arctan(\sin\theta)$.
In this ideal scenario, Alice’s measurements combined with Bob’s measurements $\bobobs{0}$ and $\bobobs{1}$ are used to perform self-testing of the shared quantum state, while Alice’s measurements combined with Bob’s $\bobobs{2}$ are utilized to extract device-independent randomness.

On one hand, the self-testing is based on the maximal violation of the tilted CHSH inequality \cite{Bamps2015}:
\begin{equation}
\label{eq:tilted_CHSH}
\begin{split}
    \expval{I_{\alpha}} = \alpha \expval{\aliceobs{3}} + \expval{\aliceobs{3}\bobobs{0}} + \expval{\aliceobs{3}\bobobs{1}} &+ \expval{\aliceobs{1}\bobobs{0}} - \expval{\aliceobs{1}\bobobs{1}} \\ 
    \expval{I_{\alpha}} \overset{\mathcal{Q}}{\leq} \sqrt{8 + 2\alpha^2} &\equiv I_\alpha^\mathcal{Q} \,,
\end{split}
\end{equation}
where $\alpha = 2 \cos\theta/\sqrt{1 + \sin^2\theta}$.
Note that the quantum saturation of this inequality does not uniquely distinguish the states defined in \eqref{eq:state_Bloch} over the entire interval $]0,\pi[$, but it only does so for $]0,\frac{\pi}{2}]$ \cite{Bamps2015}. 
This limitation arises because the tilted CHSH inequality is symmetric under the transformation $\theta \rightarrow \pi/2 - \theta$. 
By also fixing the value of $\expval{\aliceobs{3}}$ as constraint, the limitation can be avoided.

On the other hand, the randomness certification is quantified from the mean value of the operator $\mathcal{S}$ introduced in Sec.\ \ref{sec:3_out}:
\begin{equation}
\label{eq:meanTsirel}
\begin{split}
    \expval{\mathcal{S}} = &\frac{\sqrt{3}\sin{\theta}}{2-\cos{\theta}}\expval{A_1\otimes(\Pi_1-\Pi_2)}
    + \expval{A_3 \otimes \Pi_0} \\
    &+ \frac{1-2\cos{\theta}}{\cos{\theta}-2}\expval{A_3 \otimes (\Pi_1+\Pi_2)}\,,
\end{split}
\end{equation}
which, in the ideal case $\expval{\mathcal{S}}=1$, leads to the certification of $H = H_{min} = -\log_2\qty(\frac{1}{3})\approx 1.6$\ bits for $\theta = \frac{\pi}{2}$ (maximally entangled state).

We anticipate that, even when the observed correlations deviate from the ideal case, where $\expval{I_\alpha} = I_\alpha^{\mathcal{Q}}$ and $\expval{\mathcal{S}} = 1$, it is still possible to certify nonzero randomness using numerical methods.

\subsection{Experimental setup}
\label{sec:exp_setup}
\begin{figure*}
    \centering
    \includegraphics[width=\linewidth]{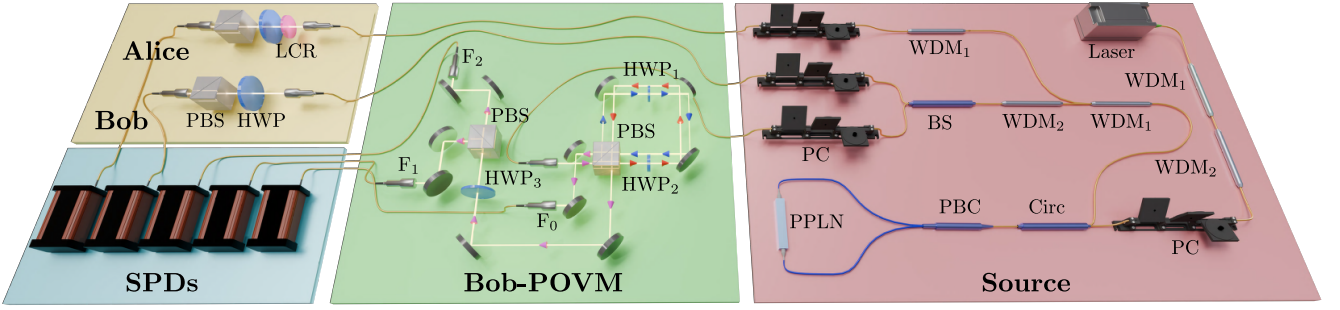}
    \caption{Schematic of the experimental setup. The main components are denoted as: Wavelength Division Multiplexing (WDM), Polarization Beam Splitter (PBS), Half-Wave Plate (HWP), Beam Splitter (BS), Polarization Controller (PC), Polarization Beam Combiner (PBC), Periodically Poled Lithium Niobate (PPLN), Circulator (Circ), Liquid Crystal Retarder (LCR), and Single Photon Detector (SPD). The outputs of the POVM are denoted as $F_{0}$, $F_{1}$, and $F_{2}$. We used a fiber-based entangled photon source pumped at 1559.44 nm exploiting the cascade SHG/SPDC in a PPLN type-0 waveguide. The non-degenerate entangled photons are separated using WDM filters centered at 1553.33 nm ($\text{WDM}_1$) and 1565.55 nm ($\text{WDM}_2$). On the left, the detection system is shown, including Alice's and Bob's projective measurement setup (yellow) and Bob's POVM setup (green). The single photon detectors are SNSPDs ID281. A more detailed discussion of the experiment can be found in Appendix \ref{app:experiment}.}
    \label{fig:experimental_setup}
\end{figure*}

The experimental setup, illustrated in Fig.\ \ref{fig:experimental_setup}, is inspired by \cite{Arahira:11,PhysRevLett.117.260401}.  
We employed a fiber-based polarization-entangled photon source in a Sagnac-loop configuration, based on a periodically poled lithium niobate (PPLN) waveguide.  
The generated photon pairs are separated according to their wavelengths and distributed to Alice and Bob by means of wavelength division multiplexers (WDM), and the parameter $\theta$ of the quantum state is controlled by a polarization controller (PC) placed before the Sagnac-loop.
Alice performs standard projective polarization measurements ($A_1$ and $A_3$) with half-wave plates (HWPs) and polarizing beam splitters (PBSs), with an additional liquid crystal retarder (LCR) to compensate the relative phase between the two components of the entangled state.  
Bob, instead, can choose between standard projective measurements ($B_0$ and $B_1$) or the triangular POVM ($\mathcal{B}_2$). 
This choice is passively implemented with a 50:50 beam splitter (BS).  
The 3-outcome POVM is realized with a non-collinear Sagnac interferometer that acts as a partially polarizing beam splitter \cite{PhysRevLett.117.260401}. 
One of its output ports corresponds to the component $F_0$, while the photons exiting from the other port are directed to a projective measurement station, implemented with an HWP ($\text{HWP}_3$) and a PBS. 
The output ports of such PBS correspond to the POVM components $F_1$ and $F_2$. 
While the optical components of the POVM are fixed, the HWPs in the projective measurement stations of Alice and Bob (orage zone in Fig. \ \ref{fig:experimental_setup}) allows to choose among the measurement configurations, i.e. the measured observable.
The acquisition time of each configuration is chosen so as to collect a number of coincidence counts ranging from $10^3$ to $10^5$.  
Single-photon detectors and a timetagger record the photon arrival times, and the coincidences between Alice and Bob are used to estimate the experimental probabilities for all measurement configurations. 
Further details of the experimental setup are provided in Appendix\ \ref{app:experiment}.   

\subsection{Analysis and results}
\label{sec:exp_results}
For estimating the certifiable min-entropy from the observed (non-ideal) correlations, we employ semidefinite programming, using the NPA hierarchy \cite{navascues_bounding_2007, navascues_convergent_2008}.
In this way, we can estimate the Eve's guessing probability and conditional von Neumann entropy by means of numerical optimizations \cite{Brown2024}.
A detailed description of such optimizations is provided in Appendix \ref{sec:randomness_estimation_app}.

As anticipated, the randomness estimation is based only on the experimental mean values of the tilted CHSH \eqref{eq:tilted_CHSH}, of the observable $\aliceobs{3}$, and of the operator $\mathcal{S}$ \eqref{eq:meanTsirel}.
In principle, the parameter $\theta$ appearing in \eqref{eq:tilted_CHSH} and \eqref{eq:meanTsirel} should be chosen to match the angle set for the experimental state, $\ket{\phi_\theta}$. This choice would enable the maximum violation of \eqref{eq:tilted_CHSH}, thereby ensuring that \eqref{eq:meanTsirel} is a proper quantum boundary.
However, when experimental noise is present, this could not be the case. Moreover, in a DI perspective, the state in which the system is prepared it's not a known information.
Therefore, in order to maximize the randomness certification, the $\theta$ parameter is chosen so that the distance $\abs*{\expval{I_\alpha}_{exp} - I_\alpha^\mathcal{Q}}$ is minimized.

Although the experimental entropy estimation is solely based on the observed correlations, we also employ a simple model to predict the potential range of achievable entropy, taking into account experimental imperfections. To this end, we express the state produced by the entangled source as
\begin{equation}
\begin{aligned}
\label{eq:exp_state}
\rho_{AB} &= (1-p-c)\dyad{\phi_\theta} 
+ p \frac{\openone}{4} \\
&\quad + c \left[ \cos^2{\left( \frac{\theta}{2} \right)} \dyad{00} + \sin^2{\left( \frac{\theta}{2} \right)} \dyad{11} \right]\,,
\end{aligned}
\end{equation}
where $p \in [0,1]$ accounts for depolarization due to mixing with random noise and $c \in [0,1]$ represents decoherence that reduces the off-diagonal elements of the density matrix. 
This decoherence can arise, for instance, from alignment inaccuracies, which increase the distinguishability between the two photons in each entangled pair. 
The parameters $p$ and $c$ can be experimentally estimated from simple visibility measurements in the $\mathcal{Z}$ and $\mathcal{X}$ bases \cite{foletto2021experimental}:
\begin{equation}
\label{eq:visibPC}
\begin{split}
    V_{\mathcal{Z}} &= \Tr[\sigma_3 \otimes \sigma_3 \rho_{AB}] = 1-p\,, \\
    V_{\mathcal{X}} &= \Tr[\sigma_1 \otimes \sigma_1 \rho_{AB}] = (1-p-c) \sin{\theta}\,.
\end{split}
\end{equation}
The model’s prediction is then obtained by calculating the correlations when ideal measurements are performed on the state in Eq.~\eqref{eq:exp_state}. 
The aim of this model is to provide a rough estimate of the achievable entropies by performing simple measurements on the state produced by the entangled source, without requiring the full construction of the experimental setup. While this model neglects measurement non-idealities, which a more refined model would account for, we show that its predictions are consistent with the observed values.

\begin{figure*} 
 \centering 
    \includegraphics[width=\linewidth]{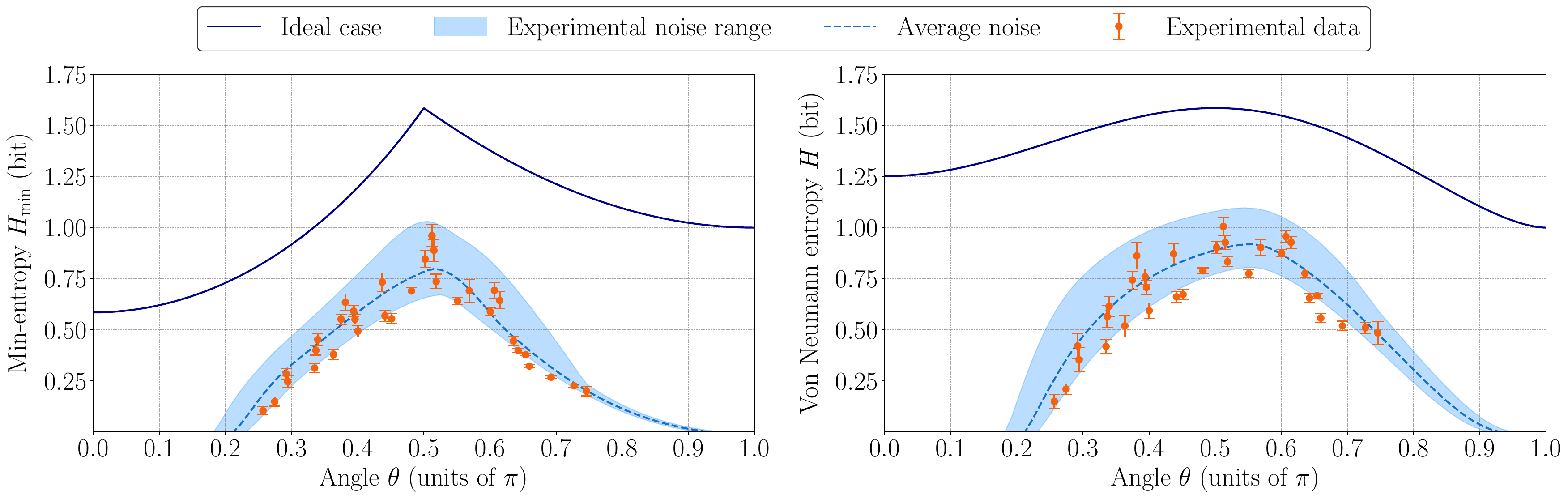} 
    \caption{Min-entropy $H_{\mathrm{min}}$ (left) and von Neumann entropy $H$ (right) as functions of the angle $\theta$. 
    Orange points represent experimental data, with error bars showing standard deviations from 100 Monte Carlo simulations.
    The light blue-shaded area indicates the range of entropy values from numerical simulations, considering one standard deviation around the experimental mean parameters $p$ and $c$.
    The dashed light blue curves represent numerical simulations using the average values of $p$ and $c$.
    The blue curves correspond to the analytical entropies.}
    \label{fig:exp_results}
\end{figure*}

We collected data for 33 measurement configurations (distinct values of $\theta$) of the experiment, obtaining an average of \(99.0\%\) of the tilted CHSH value relative to its maximum quantum bound, with a standard deviation of \(0.6\%\). 
For the operator \(\mathcal{S}\), we measured an average value of \(0.94\) with a standard deviation of \(0.03\). 
All data are presented in Table \ref{tab:exp_results_appendix} in Appendix \ref{app:experiment}.
The entropies estimated from the experimental data are shown as orange points in Fig. \ref{fig:exp_results}.
By calculating the average noise contribution and its standard deviation through \eqref{eq:visibPC}, and by considering the state model \eqref{eq:exp_state}, we estimated the range of achievable entropy, which is shown as a light blue region in the figure.
The dashed line inside this region represents the average noise case, i.e., $p = 0.007$ with a standard deviation of $0.004$, and $c = 0.01$ with a standard deviation of $0.01$.

It is worth noting that the obtained von Neumann entropy is always higher than the corresponding min-entropy, as expected.
However, the difference between the two changes as the tilting of the state deviates from the maximally entangled state. 
In the latter case ($\theta \approx \pi /2$), the ratio obtained between the experimentally von Neumann entropy and min-entropy is about 1.07, indicating that the two entropies are similar. 
However, as the state becomes increasingly tilted, this ratio grows significantly. 
In particular, we found a ratio of around 1.44 for $\theta \approx 0.26 \pi$ and 2.43 for $\theta \approx 0.76 \pi$.

In conclusion, we conducted a proof-of-principle experiment to extract device-independent randomness by imposing the tilted CHSH inequality and the Tsirelson bound in a real-world scenario, demonstrating feasibility under experimental conditions such as noise, imperfect alignment, and tilted entangled states. 
We certified randomness from entangled states with tilting angles up to $\theta \approx 0.26 \ \pi$ and $\theta \approx 0.76 \ \pi$, showing the effectiveness of the method even for highly tilted states. 
The maximum obtained min-entropy is $0.96 \pm 0.05$, while the maximum von Neumann entropy reached is $1.01 \pm 0.04$, estimated with a $99.8\pm 0.2\%$ violation of the tilted CHSH inequality and a value of $\expval{\mathcal{S}}=0.953\pm 0.001$.

\section{Conclusions}
\label{sec:Conclusions}
In this work, we have explored the correlations that arise from rank-one qubit measurements with a spatially separated party. 
Our central finding is that for any shared entangled state and any rank-one POVM, it is possible to construct a protocol such that the joint correlations lie on the boundary of the quantum set and we expressed this boundary in terms of a Tsirelson inequality.
A particularly significant implication of our results is that, when the correlations originate from extremal POVMs, the saturation of such an inequality guarantees the security of the probability distribution. 
Specifically, in such cases, correlations with a potential eavesdropper factorize, meaning the guessing probability and the conditional von Neumann entropy match those expected in a fully trusted-device scenario. 
Furthermore, we provided insight into how this method could extend to correlations generated by non-extremal POVMs. 
Our Tsirelson inequality enables randomness certification that relies not on the full joint correlations between the two separated parties but solely on the experimental value of the operator defining the inequality.
This significantly simplifies numerical certification via semidefinite programming, as it requires fewer constraints.
We validated our approach with a proof-of-concept experiment, further demonstrating its robustness in real-world scenarios.

The results of this paper could be useful in both practical applications and comprehension of fundamental aspects of quantum correlations. 
On the applied side, they are directly relevant to the security analysis of device-independent protocols, including randomness extraction and cryptographic tasks. 
They could also extend to certain semi-device-independent scenarios, where the unitary measurement operators and the shared quantum state are assumed or obtained through alternative methods, rather than being derived from the violation of tilted CHSH inequalities.
On a more fundamental level, our results provide deeper insight into the relationship between device-independent probability distribution and randomness, while also enhancing our understanding of the geometric structure of the quantum correlation and the POVM set.
Moreover, one important aspect we did not address is the possibility of self-testing the POVMs. 
Prior work \cite{Woodhead_2020} demonstrated self-testing for extremal POVMs under a full-rank assumption. 
A natural question arises: what conclusions can be drawn if this assumption is relaxed? 
In this sense, our findings provide a foundation for further investigations, such as those initiated in \cite{baptista2023}.
Finally, the connection between randomness extraction and the saturation of the Tsirelson bound makes our results particularly relevant for noise analysis. 
If the Tsirelson bound is not maximally saturated, how much randomness can be extracted? Can we derive an analytical bound linking randomness to the guessing probability when noise is introduced in both the Tsirelson bound and the tilted CHSH inequality used to certify Alice's operator, similar to the approach in \cite{Pironio2010}? 
Moreover, although we considered qubit POVMs correlations, our method may be extended for studying qudit POVMs correlations, such as the one studied in \cite{Borkala2022, sarkar2023, sarkar2024,farkas2024maximaldeviceindependentrandomnessdimension}.
These questions remain open for future research and further refinements.

\section*{Acknowledgements}
This work was supported by European Union's Horizon Europe research and innovation program under the project Quantum Secure Networks Partnership (QSNP), grant agreement No 101114043. Views and opinions expressed are however those of the authors only and do not necessarily reflect those of the European Union or European Commission-EU. Neither the European Union nor the granting authority can be held responsible for them.

\section*{Author contributions}
L.C. and M.P. addressed the theoretical proofs with the help of A.P. and M.S, and the supervision of G.V.. 
A.P. and M.S. carried out the experiment with the supervision of D.G.M..
M.S., M.P., A.P and L.C. analyzed the experimental data. 
A.P. and M.A. realized the entangled photon source used in the experiment. 
G.V. and P.V. supervised the entire work.
All authors participated in the discussion of results and contributed to the final manuscript.

\bibliography{biblio}

\onecolumngrid
\appendix

\section{Derivation of Tsirelson bound}
\label{appendix:coefficients}
In this appendix, we complement the discussion in Sec.\ \ref{sec:Methods} by showing how to find the Tsirelson boundary associated with a given rank-one POVM $\{ F_b \} $ on Bob's side, acting on the state
\begin{equation}
    \ket{\phi_{\theta}}=\cos \qty(\frac{\theta}{2}) \ket{00}+\sin \qty(\frac{\theta}{2}) \ket{11} 
\end{equation}
entangled with Alice, with $\theta =]0,\pi [$
As discussed in Sec.\ \ref{sec:Methods}, in order to find a Tsirelson bound we need to find a combination $\tilde{\Lambda}_b$ of Alice's operators such that 
\begin{align} 
\label{eq:app_cond_op}
    \expval{\tilde{\Lambda}_b \otimes F_b}{\phi_\theta}&=0\,,\\
     \tilde{\Lambda}_b^2\ket{\phi_\theta}&=\tilde{\Lambda}_b\ket{\phi_\theta} \ .
\end{align}
The second condition means that we want our operator to be a projector when acting on the state, so we start looking for an operator written as $ \tilde{\Lambda}_b = \ketbra{\alpha_b^\perp}$ with
\begin{equation}
\label{eq:lambda_bloch}
    \ket{\alpha_b^\perp}=\cos \left(\frac{\theta_b^A}{2}\right) \ket{0}+e^{i \phi_b^A}\sin \left(\frac{\theta_b^A}{2}\right) \ket{1} \ .
\end{equation}
To solve \eqref{eq:app_cond_op}, 
instead, we note that, being of rank-one, each element $F_b$ of the POVM  can be written in terms of vectors $\ket{\beta_b}$ as 
\begin{equation}
F_b=k_b\ketbra{\beta_b}
\end{equation}
with
\begin{equation}
\ket{\beta_b}=\cos(\frac{\theta^B_b}{2})\ket0+
e^{i\phi^B_b}\sin(\frac{\theta^B_b}{2})\ket1    \ . 
\end{equation}
Hence, when $F_b$ acts on the entangled state $\ket{\phi_{\theta}}$ causes the state to collapse into a separable state
\begin{equation}
    F_b\ket{\phi_\theta}=k_b\ket{\alpha_b}\ket{\beta_b}
\end{equation}
with
\begin{equation}
\label{eq:Bloch_alpha_ort}
    \ket{\alpha_b}=\cos \left(\frac{\theta_b^B}{2}\right)\cos \left(\frac{\theta}{2}\right) \ket{0}+e^{-i\phi_b^B}\sin\left( \frac{\theta_b^B}{2}\right)\sin \left(\frac{\theta}{2}\right) \ket{1} \ .
\end{equation}
Then the mean value in \eqref{eq:app_cond_op} is zero by requiring that $\braket{\alpha_b^\perp}{\alpha_b}=0$: The choice
\begin{equation}
\begin{split}
\label{eq:sol_scal_prod}
    \phi^A_b&=-\phi^B_b \ , \\
    \tan\left(\theta^A_b/2\right)&=-\cot\left(\theta^B_b/2\right)\cot\left(\theta/2\right) 
\end{split}
\end{equation}
satisfies this requirement.
We now want to interpret this solution as a combination of Alice's operators. To do so,
we recall that the projector $\tilde{\Lambda}_b$ can also be written in terms of Pauli matrices as
\begin{equation}
    \tilde{\Lambda}_b=\frac{\openone-\sum_{j=1}^3 n_{jb}\sigma_j}{2}
\end{equation}
and the Bloch representation \eqref{eq:lambda_bloch} can be recovered with the change of coordinates
\begin{equation}
\label{eq:Bloch_to_pauli}
\begin{split}
    n_{b 1} &= -\sin \theta^A_b\cos\phi^A_b \ , \\
    n_{b 2} &= -\sin \theta^A_b\sin\phi^A_b \ , \\
    n_{b 3} &= -\cos \theta^A_b \ .
\end{split}
\end{equation}
Our assumption is that Alice is performing measurements unitarily equivalent to the Pauli matrices. 
By maximally violating an appropriate set of Bell inequalities, this can be certified in a DI scenario, up to some ambiguity in the sign of $\sigma_2$. 
Without loss of generality, we can assume that the correlations are actually generated using $+\sigma_2$: different choices would change the coefficient of $n_{b2}$ accordingly. 
Since we look for an expression in a DI scenario where Alice's observables are self-tested to be the Pauli matrices, we do the replacement
\begin{equation}
    \sigma_j \quad \rightarrow \quad A_j\,,
\end{equation}
and therefore
\begin{equation}
\label{eq:comb_A:app}
    \tilde{\Lambda}_b= \frac12(\openone-\vec{n}_{b}\cdot \vec{\sigma}) \quad \rightarrow \quad \Lambda_b= \frac12(\openone-\vec{n}_{b}\cdot \vec{A})\,,\qquad |\vec{n}_b|^2=1
\end{equation}
with the coefficients 
\begin{equation}
\label{eq:final_coeff_app}
\begin{split}
    n_{b 1}&=\frac{\sin \theta \sin \theta^B_{b}}{1+\cos \theta \cos \theta^B_{b}}\cos\phi^B_{b} \ , \\
    n_{b 2}&=-\frac{\sin \theta \sin \theta^B_{b}}{1+\cos \theta \cos \theta^B_{b}}\sin \phi^B_{b} \ , \\
    n_{b 3}&=\frac{\cos \theta+\cos \theta^B_{b}}{1+\cos \theta \cos \theta^B_{b}} \ .
\end{split}
\end{equation}
found by substituting \eqref{eq:sol_scal_prod} into \eqref{eq:Bloch_to_pauli}. In this step it can be useful to recall the trigonometric relations
\begin{equation}
    \sin \left(2 \arctan{x} \right)=\frac{2x}{1+x^2} \ , \qquad  \cos \left(2 \arctan{x} \right)=\frac{1-x^2}{1+x^2} \ .
\end{equation}
\subsection{Linear independence and extremal POVMs}
\label{appendix:lin_ind}
In this section we discuss in more details the relation between extremal POVMs and invertibility of relations \eqref{eq:PitoB}. We want to show that the change of variables \eqref{eq:PitoB} from $\{ \Pi_b \}$ to $\{ \openone, M_j \}$ is invertible if and only if the POVM $F_b$ related to $\Pi_b$ via \eqref{eq:constr_mean_value_0} and \eqref{eq:constr_mean_value}, is extremal.

We start noting that change of variables relations is invertible iff the matrix whose columns are made of vectors $\{(1,\vec{n}_b) \}_b$ admits a left inverse.
This is equivalent to requiring that the vectors $ \{ (1, \vec{n}_b) \}_b$, are linear independent.
Decomposing Bob's POVM as
\begin{equation}
    F_b=k_b(\openone+\vec{m}_b\cdot \vec{\sigma}) \ ,
\end{equation}
and using the results of the previous section, we can write these vectors in terms of the parameters $\vec{m}_b=(m_{b1},m_{b2},m_{b3})$ as
\begin{equation}
\label{eq:transf_m_to_n}
\begin{split}
    n_{1b}&=-\frac{m^B_{b1}}{1-m_{b3}^B \cos \theta}\sin \theta \ , \\
    n_{2b}&=\frac{m^B_{b2}}{1-m_{b3}^B \cos \theta}\sin \theta \ , \\
    n_{3b}&=\frac{\cos\theta-m^B_{b3}}{1-m_{b3}^B\cos \theta} \ .
\end{split}
\end{equation}
As pointed out in \cite{D'Ariano_2005}, Corollary 5, a rank-one POVM is extremal if and only if its elements are linearly independent, namely if the vectors ${(1,\vec{m}_b)}_b$ are linearly independent.
Therefore, in order to prove that \eqref{eq:PitoB} is invertible if and only if the POVM is extremal, we need to check that the transformation \eqref{eq:transf_m_to_n} does not affect linear independence.

To see this, we decompose \eqref{eq:transf_m_to_n} into two transformations. We first apply an invertible linear transformation, which preserves linear independence:
\begin{equation}
    \begin{pmatrix}
        1 & 0 & 0 & -\cos \theta \\
        0 & -\sin \theta & 0 & 0 \\
        0 & 0 & \sin \theta & 0 \\
        \cos \theta & 0 & 0 & -1  
    \end{pmatrix}
    \begin{pmatrix}
    1 \\
    m_{b1}^B \\ 
    m_{b2}^B \\ 
   m_{b3}^B
    \end{pmatrix}=
 \begin{pmatrix}
    1-m_{b3}^B\cos \theta \\
    -m_{b1}^B \sin \theta \\ 
    m_{b2}^B  \sin \theta\\ 
    \cos \theta-m_{b3}^B
    \end{pmatrix}
\end{equation}
Then, for each $b$ we rescale the vectors we just found by $1/(1-m_{b3}^B \cos \theta)$ finding 
\begin{equation}
   \frac{1}{1-m_{b3}\cos \theta}
    \begin{pmatrix}
    1-m_{b3}^B\cos \theta \\
    -m_{b1}^B \sin \theta \\ 
    m_{b2}^B  \sin \theta\\ 
    \cos \theta-m_{b3}^B
    \end{pmatrix}=
    \begin{pmatrix}
    1 \\
    n_{b1}^B \\ 
    n_{b2}^B \\ 
    n_{b3}^B
    \end{pmatrix}
\end{equation}
Being a rescaling, also this last transformation do not change linear independence.
This concludes the proof.

\section{Min-entropy for 4-outcome non-extremal POVM}
\label{sec:app_eve_strat}
In this appendix, we provide more details on the  computation of the min-entropy when the correlations are produced by the  ``square" POVM (see Sec.\ \ref{sec:non_ext_4})
\begin{equation}
\begin{split}
    F_0&=\frac{1}{2}\dyad{+} \ ,  \\
    F_1&=\frac{1}{2}\dyad{-} \ ,  \\
    F_2&=\frac{1}{2}\dyad{0} \ , \\
    F_3&=\frac{1}{2}\dyad{1} \ ,
\end{split}
\end{equation}
which can be uniquely decomposed in terms of two extremal POVMs $\mathbf{C}$ and $\mathbf{D}$:
\begin{equation}
\label{eq:decompQuadrato_app}
    \mathbf{F} \equiv \mqty(F_0 \\ F_1 \\ F_2 \\ F_3) = \frac{1}{2}\mqty(\dyad{+}\\ \dyad{-}\\ 0 \\0) + \frac{1}{2}\mqty( 0 \\0 \\ \dyad{0}\\ \dyad{1}) \equiv \frac{1}{2}\mathbf{C} + \frac{1}{2}\mathbf{D} \ .
\end{equation}

The guessing probability is defined as
\begin{equation}
\label{eq:pguess_def_app}
    p_g = \max_{E}\sum_b \expval{\Pi_b E_b}\,,
\end{equation}
where the maximization is performed over all the possible Eve's strategies compatible with the experimental data.
We define the two projective operators
\begin{equation}
    M_0^{(1)} \equiv \Pi_0+\Pi_1 \ ,  \qquad  
     M_0^{(2)} \equiv \Pi_2+\Pi_3 = \openone - M_0^{(1)}
\end{equation}
and we use them for writing the projectors $\Pi_b$ as
\begin{align}
    \Pi_0& =\frac{M_0^{(1)}\sin \theta +M_1}{2 \sin \theta} \ , \\ 
    \Pi_1&= \frac{M_0^{(1)}\sin \theta -M_1}{2 \sin \theta}  \ , \\
    \Pi_2& =\frac{M_0^{(2)}+M_3-\cos\theta M_0^{(1)}}{2} \ ,  \\
    \Pi_3&= \frac{M_0^{(2)}-M_3+\cos \theta M_0^{(1)}}{2} \ .
\end{align}
In this way, the terms inside the guessing probability become the following
\begin{align}
    \expval*{\Pi_0 E_0}&=\frac{1}{2}\expval*{M_{0}^{(1)} E_0}+\frac{1}{2 \sin \theta}\expval{M_1} \expval*{E_0} \ , \\
    \expval*{\Pi_1 E_1} &= \frac{1}{2}\expval*{M_{0}^{(1)} E_1}-\frac{1}{2 \sin \theta}\expval*{M_1} \expval*{E_1} \ , \\
    \expval*{\Pi_2 E_2} &= \frac{1}{2}\expval*{(M_{0}^{(2)}-\cos \theta M_0^{(1)})E_2}+\frac{1}{2}\expval*{M_3} \expval{E_2} \ , \\
    \expval*{\Pi_3 E_3} &=\frac{1}{2}\expval*{(M_{0}^{(2)}+\cos \theta M_0^{(1)})E_3}-\frac{1}{2}\expval*{M_3} \expval{E_3} \,.
\end{align}
Recalling that 

\begin{equation}
  \expval{M_1}=\expval{A_1}=\expval{\sigma_x\otimes \openone}{\phi_\theta}=0 \ , \qquad \expval{M_3}=\expval{A_3}=\expval{\sigma_z\otimes \openone}{\phi_\theta}=\cos \theta
\end{equation}
we simplify the expression to
\begin{equation}
    p_{\text{g}}= \max_{\text{E}}\left[\frac{\expval*{M_0^{(1)} (E_0+E_1)}}{2}+\frac{\expval*{M_0^{(2)} (E_2+E_3)}}{2}  +\cos \theta \left(\frac{\expval*{M_0^{(1)}(E_3-E_2)}}{2}-\frac{\expval*{E_3-E_2}}{2}\right) \right] \ .
\end{equation}
To upper bound this guessing probability, we note that 
\begin{equation}
\cos\theta\frac{\expval*{M_0^{(1)} (E_3-E_2)}}{2} \le
\abs{\cos\theta}\frac{\expval*{M_0^{(1)} (E_2+E_3)}}{2}\le \frac{\expval*{M_0^{(1)} (E_2+E_3)}}{2} 
\end{equation}
and 
\begin{equation}
    \frac{\expval*{M_0^{(i)}\sum_e E_e}}{2}\le \frac{\expval*{M_0^{(i)}}}{2} \ .
\end{equation}
Therefore, we conclude
\begin{equation}
\begin{split}
\label{eq:pguess_quadrato_bound_app}
    p_g &\le \frac{\expval*{M_0^{(1)}}}{2}+\frac{\expval*{M_0^{(2)}}}{2}-\cos \theta \frac{\expval*{E_3-E_2}}{2} \\
      & \le \frac{1}{2}+\frac{\abs{\cos \theta}}{2} \ .
\end{split}
\end{equation}
One can check that the previous inequality is actually an equality by constructing an explicit strategy for the eavesdropper, as done in Appendix \ref{sec:app_evestrate}.

\subsection{Optimal Eve's strategy}
\label{sec:app_evestrate}
To provide Eve's strategy that saturates the upper bound in \eqref{eq:pguess_quadrato_bound_app}, we draw inspiration from the decomposition \eqref{eq:decompQuadrato_app}.
The decomposition shows that the correlations generated by the square POVM $\mathbf{F}$ can always be obtained by Statistically choosing between the two POVMs $\mathbf{C}$ and $\mathbf{D}$, which produce correlations satisfying the boundary condition
\begin{equation}
\label{eq:bound_condition_state_app}
 \sum_{j=1}^3 A_j \otimes M_j\ket{\psi}=\ket{\psi} \,,
\end{equation}
with 
\begin{equation}
\label{eq:rel_4out_app}
    \mqty(\openone \\ M_1 \\ M_2 \\ M_3)= \mqty(1 & 1 & 1 & 1 \\ 
          \sin \theta & -\sin \theta & 0 & 0 \\ 0 & 0 & 0 & 0 \\ \cos \theta & \cos \theta & 1 & -1 ) \mqty(F_0 \\ F_1 \\ F_2 \\ F_3)\,.
\end{equation}
We recall that for calculating the upper bound \eqref{eq:pguess_quadrato_bound_app} we imposed only \eqref{eq:bound_condition_state_app}.
Geometrically, we are giving Eve the freedom to move along the boundary of the quantum correlations defined by \eqref{eq:bound_condition_state_app}, and to choose the best strategy for her.

It turns out that the upper-bound of the guessing probability \eqref{eq:pguess_quadrato_bound_app} is obtained when the POVM $\mathbf{D}$ is implemented, i.e., when Eve prepares the ideal state $\ket{\psi}=\ket{\phi_\theta}_{AB}$, Bob's device is such that
\begin{align}
    \Pi_0 &= 0 \ , \\
    \Pi_1 &= 0 \ , \\
    \Pi_2 &= \dyad{0} \ , \\
    \Pi_3 &= \dyad{1}
\end{align}
and Eve bets on the most probable outcome, which is $b=2$ when $\cos\theta > 0$, or $b=3$ when $\cos\theta < 0$.
Therefore, for this strategy we conclude that
\begin{equation}
    p_g = \max_b \expval{\Pi_b}{\phi_\theta} = \frac{1}{2} + \frac{\abs{\cos \theta}}{2}\,.
\end{equation}

Instead, if all the correlations between Alice and Bob are considered, and not just the boundary \eqref{eq:bound_condition_state_app}, then the above strategy is no longer valid.
It can be shown numerically that, in this case, the guessing probability becomes
\begin{equation}
\label{eq:strat_full_corr}
    p_g = \frac{1}{2} + \frac{\abs{\cos \theta}}{4}\,.
\end{equation}
We verified the above guessing probability for some values of $\theta$ with the numerical methods reviewed in Appendix \ref{sec:randomness_estimation_app}, using an NPA order of 2 with the solver SPDA-DD \cite{nakata2010numerical} and setting a solver precision of $10^{-12}$.
A simple strategy achieving the expression in \eqref{eq:strat_full_corr} is the one with
\begin{align}
    \Pi_0 &= \dyad{+}\otimes\dyad{0} \ , \\
    \Pi_1 &= \dyad{-}\otimes\dyad{0} \ , \\
    \Pi_2 &= \dyad{0}\otimes\dyad{1}\ , \\
    \Pi_3 &= \dyad{1}\otimes\dyad{1} \,,
\end{align}
where the projectors act on an Hilbert space $\mathcal{H}_B \otimes \mathcal{H}_{B'}$,
the joint state is $\ket{\psi} = \ket{\phi_\theta}_{AB}\ket{\phi^+}_{B'E}$, and Eve measures the POVM
\begin{align}
    \qty{E_0, E_1, E_2, E_3} &= \qty{\dyad{0}, 0, \chi\dyad{1}, (1-\chi)\dyad{1}} \ ,\\
    \chi &= \begin{cases}
        1 & \cos\theta > 0 \ , \\
        0 & \cos\theta < 0 \,.
    \end{cases}
\end{align}

\section{Experiment}
\label{app:experiment}
To verify the feasibility of randomness extraction based on the violation of the Tsirelson inequalities introduced in the main text, we conducted a proof-of-principle experiment using an optical setup, which we describe in this appendix.

\subsection{Protocol description}
\label{appendix:protocol_description}
We consider a scenario in which Alice has two possible inputs $x = 1,3$, while Bob has three possible inputs $y = 0,1,2$.
The possible outcomes of Alice are $a = 0,1$, the possible outcomes of Bob are $b=0,1$ for $y=0,1$ and $b=0,1,2$ for $y=2$.

The shared entangled state is
\begin{equation}
\label{eq:app_ent_state}
    \ket{\phi_{\theta}} = \cos(\frac{\theta}{2})\ket{00} + \sin(\frac{\theta}{2}) \ket{11}\,,
\end{equation}
with $\theta \in ]0,\pi[$, and Alice's ideal measurements are the observables
\begin{equation}
    \aliceobs{1} = \sigma_1 \ , \qquad \aliceobs{3} = \sigma_3\,.
\end{equation}
On the other hand, Bob's ideal measurements are
\begin{equation}
    \bobobs{0} = \cos\mu \sigma_3 + \sin\mu \sigma_1 \ ,  \qquad \bobobs{1} = \cos\mu \sigma_3 - \sin\mu \sigma_1 \ , \qquad \bobobs{2} = \qty{F_b}_{b=0,1,2} \,,
\end{equation}
with $\mu = \arctan(\sin\theta)$, and the set $\qty{F_b}$ forms the triangular POVM introduced in Sec.\ \ref{sec:3_out}
\begin{align}
\label{eq:mercedesPOVM_app}
\begin{split}
    F_0 &= \frac{1}{3}\qty(\mathbb{1} + \sigma_3)=\frac{2}{3}\ketbra{0} \ , \\
    F_1 &= \frac{1}{3}\qty(\mathbb{1} + \frac{\sqrt{3}}{2}\sigma_1 - \frac{1}{2}\sigma_3)=R\qty(\frac{\pi}{3})\ketbra{0}R\qty(-\frac{\pi}{3}) \ , \\
    F_2 &= \frac{1}{3}\qty(\mathbb{1} - \frac{\sqrt{3}}{2}\sigma_1 - \frac{1}{2}\sigma_3)=R\qty(-\frac{\pi}{3})\ketbra{0}R\qty(\frac{\pi}{3})\,,
    \end{split}
\end{align}
with the rotation matrix
\begin{equation}
    R(\delta) = \mqty( \cos\delta & -\sin\delta \\ \sin\delta & \cos\delta )\,.
\end{equation}
The correlations generated by this choice saturate the Tsirelson bound of the tilted CHSH inequality
\begin{equation}
    \expval{I_{\alpha}} = \alpha \expval{\aliceobs{3}} + \expval{\aliceobs{3}\bobobs{0}} + \expval{\aliceobs{3}\bobobs{1}} + \expval{\aliceobs{1}\bobobs{0}} - \expval{\aliceobs{1}\bobobs{1}} \overset{\mathcal{Q}}{\leq} \sqrt{8 + 2\alpha^2}
\end{equation}
when $\alpha = \frac{2 \cos\theta}{\sqrt{1 + \sin^2\theta}}$. 
This saturation allows to self-test the action of the measurements $A_1,A_3$ and $B_0,B_1$ on the state in the range $\theta \in ]0,\pi/2]$ \cite{Bamps2015}.  
By also fixing the value of $\expval{\aliceobs{3}}$ as constraint, it is possible to enlarge the self-testing to the entire interval $]0,\pi[$. 
The correlations produced by Alice's operators and $B_2$, instead, saturate the Tsirelson 
\begin{equation}
\label{eq:S_protocol_appendix}
    \expval{\mathcal{S}} = \frac{\sqrt{3}\sin{\theta}}{2-\cos{\theta}}\expval{A_1\otimes(F_1-F_2)}
    + \expval{A_3 \otimes F_0} + \frac{1-2\cos{\theta}}{\cos{\theta}-2}\expval{A_3 \otimes (F_1+F_2)} = 1
\end{equation}
as discussed in the main text.

\subsection{Experimental setup with entangled photons}
\label{section:experimental_setup}
The experimental setup is illustrated in Fig.\ \ref{fig:experimental_setup}.
We used a fiber-based, non-degenerate, polarization-entangled photon source in a Sagnac loop configuration.
The source crystal is a type-0 Periodically Poled Lithium Niobate (PPLN) waveguide, provided by HC Photonics, following the scheme proposed in \cite{Arahira:11}.
Pump photons with a wavelength of $\lambda_p = 1559.44$ nm interact with the crystal through second harmonic generation (SHG), producing photons at a wavelength of $\lambda_{SHG} = 779.72$ nm.
These latter interact with the crystal via spontaneous parametric down-conversion (SPDC), generating a pair of photons (signal and idler) at wavelengths $\lambda_s= 1553.33$ nm and $\lambda_i=1565.55$ nm. 
Schematically, the conversion reads as:
\begin{equation} \ket{V,\lambda_p}\ket{V,\lambda_p}\xrightarrow{SHG}\ket{V,\lambda_{SHG}}\xrightarrow{SPDC}\ket{V,\lambda_s}\ket{V,\lambda_i} \,.
\end{equation}
A polarization controller (PC) is used to manipulate the pump state entering the Sagnac loop, which can generally be written as
\begin{equation}
\label{eq: pump state}
    \cos(\frac{\theta'}{2}) \ket{H} + e^{i\varphi} \sin(\frac{\theta'}{2}) \ket{V} \,.
\end{equation} 
At the input of the interferometer, a polarization beam combiner (PBC) splits the horizontal and vertical polarizations of the pump into two polarization-maintaining (PM) fibers, such that the light entering both sides of the PPLN waveguide matches the crystal generation axis.  

Photons generated via the SPDC process are recombined at the PBC with a polarization state of either $\ket{H, \lambda_s}\ket{H, \lambda_i}$ or $\ket{V, \lambda_s}\ket{V, \lambda_i}$, depending on the PBC port at which they combine. 
They are directed to the output of the source by a circulator and sent to Alice's and Bob's locations depending on their wavelengths.  
This separation is achieved using a cascade of $\text{WDM}_1$ and $\text{WDM}_2$, which also serve to remove any residual pump light.  
Finally, the output state is
\begin{equation}
\label{eq: entangled states}
    \ket{\psi} = \cos(\frac{\theta'}{2}) \ket{HH} + e^{i(\varphi+\varphi_{\theta'})} \sin(\frac{\theta'}{2}) \ket{VV} \,,
\end{equation}
where $\varphi_{\theta'}$ is an additional phase that may be introduced by asymmetries in the Sagnac loop. 

The output photons are directed to the measurement setup through a polarization controller to pre-compensate for fiber-induced transformations on the state.  
The projective measurement setups for Alice and Bob consists of half-wave plates (HWPs) and polarizing beam splitters (PBSs).  
Additionally, Alice's measurement station includes a liquid crystal retarder (LCR) to compensate for the relative phase $\varphi + \varphi_{\theta}$ in \eqref{eq: entangled states}, thereby ensuring the state takes the form given in \eqref{eq:app_ent_state}.  
On the other side, Bob chooses to perform either a projective measurement associated with $B_0$ and $B_1$, or the triangular POVM: this choice is made by a 50:50 fiber beam splitter (BS).
The setup of the POVM is based on a free-space, non-collinear Sagnac loop to implement a partially polarizing beam splitter, which is a device that has different splitting ratios of horizontally and vertically polarized components of the input light.
The optical scheme follows the one of \cite{PhysRevLett.117.260401}.
The incoming light is split by a PBS, directing each polarization component along separate paths, enabling polarization-dependent transformations.  
Inside the Sagnac interferometer, two HWPs are placed: $\text{HWP}_1$ in the reflected branch (to manipulate vertically polarized light) and $\text{HWP}_2$ in the transmitted branch (to manipulate horizontally polarized light).
$\text{HWP}_1$ ($\theta_1 = 0$) is set such that all vertically polarized light is directed towards the second output port of the interferometer (leading to detectors $F_1$ and $F_2$).
$\text{HWP}_2$ ($\theta_2 = \frac{1}{2}\arcsin{\sqrt{\frac{2}{3}}}$) is adjusted so that two-thirds of the horizontally polarized light exits through the first output port (towards detector $F_0$). 
Finally, at the second output port, $\text{HWP}_3$ ($\theta_3 = \frac{\pi}{8}$), together with a PBS, implements a projective measurement in the diagonal basis. In this way, it is straightforward to show (see the experimental setup in Fig.\ \ref{fig:experimental_setup}) that, for a generic qubit input state, the probabilities of detecting a photon at the outputs corresponding to ports $F_0$, $F_1$, and $F_2$ coincide with the ones given by the POVM components in Eq.\ \ref{eq:mercedesPOVM_app}.

The outputs of the measurement setup are connected to single-photon detectors (SPDs), specifically the SNSPD ID281.  
The arrival times of the photons are recorded by a time tagger with 1 ps resolution and RMS jitter of 42 ps.  
The experimental probabilities are retrieved by measuring the coincidences between the events recorded by Alice and Bob for all possible measurement combinations.  
For a typical experiment, the number of coincidences detected for each measurement ranges from $10^3$ to $10^5$, with a coincidence window of 150 ps.  

\begin{table*}
\caption{Experimental results. The number of coincidences detected for each measurement ranges from $10^3$ to $10^5$, with a coincidence window of 150 ps.}
\label{tab:exp_results_appendix}
\begin{tabular*}{\linewidth}{@{\extracolsep{\fill}}ccccc}
\toprule\toprule
$\theta$ (rad) & $\langle I_\alpha \rangle$ & $\langle \mathcal{S} \rangle$ & $H_{\mathrm{min}}$ & $H$ \\
\midrule
$0.807$ & $3.208\pm0.005$ & $0.939\pm0.001$ & $0.10\pm0.02$ & $0.15\pm0.03$ \\
$0.862$ & $3.146\pm0.004$ & $0.938\pm0.001$   & $0.15\pm0.02$ & $0.21\pm0.02$ \\
$0.917$ & $3.114\pm0.005$ & $0.919\pm0.001$   & $0.28\pm0.03$ & $0.42\pm0.06$ \\
$0.924$ & $3.106\pm0.005$ & $0.895\pm0.001$   & $0.25\pm0.03$ & $0.35\pm0.06$ \\
$1.052$  & $2.998\pm0.006$ & $0.932\pm0.002$   & $0.40\pm0.02$ & $0.56\pm0.05$ \\
$1.058$  & $2.979\pm0.006$ & $0.905\pm0.002$   & $0.31\pm0.02$ & $0.42\pm0.03$ \\
$1.067$  & $2.979\pm0.007$ & $0.931\pm0.002$   & $0.38\pm0.03$ & $0.52\pm0.05$ \\
$1.141$  & $2.952\pm0.006$ & $0.854\pm0.003$   & $0.45\pm0.03$ & $0.61\pm0.05$ \\
$1.177$  & $2.915\pm0.005$ & $0.964\pm0.001$   & $0.55\pm0.02$ & $0.74\pm0.04$ \\
$1.198$  & $2.915\pm0.006$ & $0.966\pm0.001$   & $0.63\pm0.04$ & $0.86\pm0.06$ \\
$1.238$  & $2.873\pm0.006$ & $0.964\pm0.001$   & $0.55\pm0.02$ & $0.71\pm0.03$ \\
$1.244$  & $2.878\pm0.006$ & $0.963\pm0.001$   & $0.59\pm0.03$ & $0.76\pm0.04$ \\
$1.257$  & $2.883\pm0.005$ & $0.840\pm0.001$   & $0.49\pm0.03$ & $0.59\pm0.04$ \\
$1.373$  & $2.834\pm0.009$ & $0.970\pm0.002$   & $0.73\pm0.04$ & $0.87\pm0.05$ \\
$1.386$  & $2.812\pm0.007$ & $0.911\pm0.002$   & $0.57\pm0.03$ & $0.66\pm0.03$ \\
$1.418$  & $2.775\pm0.008$ & $0.968\pm0.002$   & $0.55\pm0.02$ & $0.67\pm0.03$ \\
$1.512$  & $2.787\pm0.004$ & $0.959\pm0.001$   & $0.69\pm0.02$ & $0.79\pm0.01$ \\
$1.577$  & $2.810\pm0.005$ & $0.951\pm0.001$   & $0.85\pm0.04$ & $0.90\pm0.03$ \\
$1.609$  & $2.788\pm0.006$ & $0.965\pm0.002$   & $0.74\pm0.04$ & $0.83\pm0.02$ \\
$1.619$  & $2.819\pm0.004$ & $0.939\pm0.001$   & $0.89\pm0.05$ & $0.93\pm0.03$ \\
$1.629$  & $2.825\pm0.005$ & $0.953\pm0.001$   & $0.96\pm0.05$ & $1.01\pm0.04$ \\
$1.730$  & $2.810\pm0.005$ & $0.927\pm0.001$   & $0.64\pm0.02$ & $0.78\pm0.02$ \\
$1.787$  & $2.838\pm0.008$ & $0.963\pm0.002$   & $0.69\pm0.06$ & $0.90\pm0.04$ \\
$1.886$  & $2.876\pm0.004$ & $0.969\pm0.001$  & $0.59\pm0.02$ & $0.87\pm0.02$ \\
$1.906$  & $2.897\pm0.003$ & $0.973\pm0.001$  & $0.69\pm0.04$ & $0.96\pm0.03$ \\
$1.931$  & $2.907\pm0.003$ & $0.976\pm0.001$  & $0.64\pm0.04$ & $0.93\pm0.03$ \\
$1.997$  & $2.896\pm0.007$ & $0.966\pm0.002$   & $0.40\pm0.01$& $0.66\pm0.02$ \\
$2.019$  & $2.943\pm0.005$ & $0.976\pm0.001$   & $0.45\pm0.02$ & $0.78\pm0.02$ \\
$2.055$  & $2.948\pm0.003$ & $0.973\pm0.001$  & $0.38\pm0.01$& $0.67\pm0.01$ \\
$2.072$  & $2.952\pm0.006$ & $0.941\pm0.002$   & $0.32\pm0.01$& $0.56\pm0.02$ \\
$2.175$  & $3.039\pm0.007$ & $0.959\pm0.002$   & $0.27\pm0.01$& $0.52\pm0.02$ \\
$2.284$  & $3.176\pm0.004$ & $0.944\pm0.001$  & $0.23\pm0.01$ & $0.51\pm0.03$ \\
$2.343$  & $3.241\pm0.007$ & $0.953\pm0.002$   & $0.20\pm0.02$ & $0.49\pm0.06$ \\
\bottomrule\bottomrule
\end{tabular*}
\end{table*}

In Table\ \ref{tab:exp_results_appendix} the experimental results related to Fig.\ \ref{fig:exp_results} are shown.
The parameter $\theta'$ is fixed by the first PC placed before the Sagnac source (see \eqref{eq: pump state}), while the parameter $\theta$ is the one defining the expression of $\mathcal{S}$ (see \eqref{eq:S_protocol_appendix}).
Specifically, $\theta$ is obtained through the minimization of the distance between the experimental $\expval{I_\alpha}$ and the quantum bound $I_\alpha^\mathcal{Q}$:
\begin{equation}
\begin{split}
    &\min_{\theta} \abs{\expval{I_\alpha} - I_\alpha^\mathcal{Q}} \\
    &\text{ s.t.} \expval{I_\alpha} \leq I_\alpha^\mathcal{Q}\,.
\end{split}
\end{equation}
Finally, $\expval{A_3}$, $\expval{\mathcal{S}}$  and $\expval{S_{CHSH}}$ are the experimental values of the Alice's measurement $A_3$, of the operator $\mathcal{S}$ and the CHSH parameter 
\begin{equation}
    \expval{S_{CHSH}}_{exp} = \expval{I_{\alpha}}_{exp} - \alpha \expval{A_3}_{exp}\,.
\end{equation}
These three values are constrained in the SDP optimizations, described in the next section, estimating the min and von Neumann entropy.

\subsection{Randomness estimation}
\label{sec:randomness_estimation_app}
Given the correlations $p_{AB}^{exp}(a,b|x,y)$, we call a quantum commuting strategy a set $\mathcal{R} = \qty{\ket{\psi}, \aliceproj{x}{a}, \bobproj{y}{b}}$ such that
\begin{align}
    &\ket{\psi} \in \mathcal{H} \\
    &\aliceproj{x}{a},\bobproj{y}{b} \in \text{Proj}(\mathcal{H}) \\
    &\sum_a \aliceproj{x}{a} = \sum_b \bobproj{y}{b} = \openone \\
    &\qty[\aliceproj{x}{a},\bobproj{y}{b}] = 0 \\
    &p_{AB}^{exp}(a,b|x,y) = \expval{\aliceproj{x}{a}\bobproj{y}{b}}{\psi}\,,
\end{align}
where with $\text{Proj}(\mathcal{H})$ we indicate the set of projectors acting on $\mathcal{H}$.

The guessing probability is computed through the following noncommutative polynomial optimization problem 
\begin{equation}
\label{eq:SDP_problem}
\begin{split}
    p_g &= \sup_{\mathcal{R},E_e} \sum_{b} \expval{\Pi_b E_{b}}{\psi} \\
    \text{s.t.} \hspace{0.5cm} &\sum_b E_b = \openone \\
    &{E_e}^\dagger E_e = E_e \\
    &\sum_a \aliceproj{x}{a} = \sum_b \bobproj{y}{b} = \openone \quad \forall x,y \\
    &{\aliceproj{x}{a}}^\dagger \aliceproj{x}{a} = \aliceproj{x}{a} \quad \forall a,x\\
    &{\bobproj{y}{b}}^\dagger\bobproj{y}{b} = \bobproj{y}{b} \quad \forall b,y \\
    &\qty[\aliceproj{x}{a}, \bobproj{y}{b}] = \qty[\aliceproj{x}{a}, E_{e}] = \qty[\bobproj{y}{b}, E_{e}] = 0 \quad \forall a,b,x,y,e \\
    &\left.
        \begin{aligned}
            &\expval{S_{CHSH}} = \expval{S_{CHSH}}_{exp} \\
            &\expval{\aliceobs{3}} = \expval{\aliceobs{3}}_{exp} \\
            &\expval{\mathcal{S}} = \expval{\mathcal{S}}_{exp}
        \end{aligned}
    \right\} \til \text{Experimental constraints}
\end{split}
\end{equation}
In the above optimization, the set of projectors $\qty{E_e}$ represents all the possible quantum operations, commuting with Alice and Bob operators, that Eve can perform on the state in order to guess the Bob's outcome $b$.
The min-entropy is then calculated as $H_{min}(B|Y=2,E) = -\log_2 p_g$. We stress that randomness is certified only on Bob’s POVM input $y=2$.

The conditional von Neumann entropy is lower bounded through the following noncommutative polynomial optimization problem \cite{Brown2024}
\begin{equation}
\label{eq:VN_problem}
\begin{split}
    H(B|Y=2,E) &\geq c_m + \sum_{i=1}^{m-1}\frac{\omega_i}{t_i \ln 2} \til\inf_{\mathcal{R},Z_e} \sum_{b} \expval{\Pi_b \qty( Z_{b} + Z_b^\dagger + (1-t_i)Z_b^\dagger Z_{b}) + t_i Z_{b}Z_b^\dagger}{\psi} \\
    \text{s.t.} \hspace{0.5cm} &\sum_a \aliceproj{x}{a} = \sum_b \bobproj{y}{b} = \openone \quad \forall x,y \\
    &{\aliceproj{x}{a}}^\dagger \aliceproj{x}{a} = \aliceproj{x}{a} \quad \forall a,x\\
    &{\bobproj{y}{b}}^\dagger\bobproj{y}{b} = \bobproj{y}{b} \quad \forall b,y \\
    &\qty[\aliceproj{x}{a}, \bobproj{y}{b}] = \qty[\aliceproj{x}{a}, Z_{e}] = \qty[\bobproj{y}{b}, Z_{e}] = \qty[\aliceproj{x}{a}, Z_{e}^\dagger] = \qty[\bobproj{y}{b}, Z_{e}^\dagger] = 0 \quad \forall a,b,x,y,e \\
    &\left.
        \begin{aligned}
            &\expval{S_{CHSH}} = \expval{S_{CHSH}}_{exp} \\
            &\expval{\aliceobs{3}} = \expval{\aliceobs{3}}_{exp} \\
            &\expval{\mathcal{S}} = \expval{\mathcal{S}}_{exp}
        \end{aligned}
    \right\} \til \text{Experimental constraints}
\end{split}
\end{equation}
Here, $t_1,\dots,t_m$ and $\omega_1,\dots,\omega_m$ are the nodes and weights of an $m$-point Gauss-Radau quadrature on $[0,1]$ with endpoint $t_m = 1$, and the operators $Z_e$ can be seen as the counterpart of the projectors $E_e$ appearing in \eqref{eq:SDP_problem}.
However, in this case the operators $Z_e$ are not hermitian and therefore not projectors.
We point out that, in \eqref{eq:VN_problem}, we implemented some of the tips proposed in \cite{Brown2024} in order to get a faster computation, but in general a less tight lower bound.

The last three experimental constraints are retrievable by taking linear combinations of the correlations $p_{AB}^{exp}(a,b|x,y)$, for example, the experimental value of the CHSH is calculated as
\begin{equation}
    \expval{S_{CHSH}}_{exp} = \sum_{x,y,a,b = 0}^1 (-1)^{a+b+x\cdot y} \til p_{AB}^{exp}(a,b|x,y)\,.
\end{equation}

Both the above optimization problems are relaxed to semidefinite programs exploiting NPA hierarchy \cite{navascues_bounding_2007, navascues_convergent_2008}, setting it at order 2 for all optimizations, performed thanks to Ncpol2sdpa \cite{Wittek2015}.
For the guessing probability, we adopted the solver SDPA-DD \cite{nakata2010numerical} with a solver precision of $10^{-12}$, while for the von Neumann entropy, we adopted the solver SDPA \cite{Yamashita2010A} with a solver precision of $10^{-7}$.

\section{Randomness and detection efficiency}
In this appendix we present a simulation of the min-entropy certified with the protocol described in Section \ref{section:experimental_implementation} in presence of detection losses.
Specifically, we consider the case in which the shared state between Alice and Bob is maximally entangled ($\theta = \pi/2$), $\ket{\phi^+}$, and the two users measure the CHSH ideal measurements for the self-testing part of the protocol.

In case of no detection, we consider the strategy based on the deterministic assignment of one of the possible outcomes. 
Mathematically, given the efficiency $\eta$ (equal for all measurements) we model the measurement POVMs as
\begin{align}
    \Lambda^x_0(\eta) &= \eta\Lambda^x_0 \,,\\
    \Lambda^x_1(\eta) &= \eta\Lambda^x_1 + (1-\eta) \openone \,,
\end{align}
where $\Lambda^x_a$ are the ideal projectors of the Alice's measurements, and
\begin{align}
    \Pi^y_0 (\eta) &= \eta\Pi^y_0 \\
    \Pi^y_1(\eta) &= \eta\Pi^y_1 + (1-\eta) \openone \\
    F_0 (\eta) &= \eta F_0 \\
    F_1 (\eta) &= \eta F_1 \\
    F_2 (\eta) &= \eta F_2 + (1-\eta)\openone 
\end{align}
where $\Pi^y_b$ are the ideal projectors of the Bob's projective measurements and $F_b$ are the ideal elements of the triangular POVM.
In case of photonic implementations, this deterministic strategy requires at least $k-1$ detectors per measurements, with $k$ the number of possible measurement outcomes: For the implemented protocol, the total number is 6.

The CHSH value, $\expval{S_{Bell}}$ and POVM Tsirelson bound \eqref{eq:baptista_bound} then become
\begin{align}
    \expval{S_{Bell}(\eta)} &= 2 + 2 \eta (-2 + \eta +\sqrt{2}\eta)\,, \\
    \expval{\mathcal{S}(\eta)} &= \frac12 (1 + \sqrt{3} - 2 (1 + \sqrt{3}) \eta + (3 + \sqrt{3})\eta^2)\,.
\end{align}

\begin{figure}
    \centering
    \includegraphics[width=0.5\linewidth]{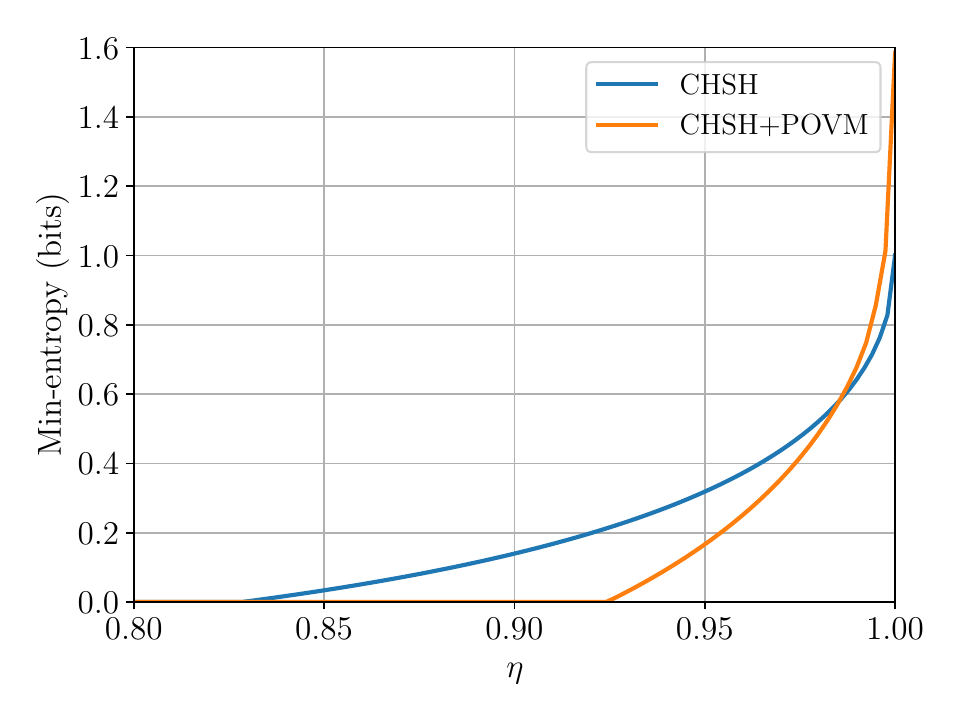}
    \caption{Min-entropy of a CHSH-based randomness protocol and of the protocol described in Section \ref{section:experimental_setup} of the main text when $\theta = \pi/2$.
    The computations are carried out with the methods of Appendix \ref{sec:randomness_estimation_app} with an NPA order 3 and solver SDPA-DD. 
    }
    \label{fig:scaneta}
\end{figure}

In Fig.\ \ref{fig:scaneta}, we show the min-entropy of the POVM outcomes when the two expressions above are constrained (orange curve), observing that the entropy drops to zero at $\eta \approx 92.5\%$.
We also compare the POVM-based protocol with a randomness generation protocol relying solely on CHSH measurements (blue curve).
The intersection of the two curves occurs at $\eta \approx 98.5\%$.

We emphasize that the maximally entangled state and CHSH measurements, which are optimal for randomness generation in the ideal case of no detection losses, may not be optimal when detection losses are present.
Indeed, in general, the shared quantum state and the measurements used for estimating $\expval{S_{Bell}}$ should be optimized as a function of the detection efficiency.
For example, one should consider a doubly-tilted CHSH inequality, rather than the standard CHSH inequality, for maximizing the nonlocality certification \cite{Gigena2025}, and consequently also the Tsirelson bound should be modified in order to maximize the entropy certification.
We let such investigations for future works.

\end{document}